\documentclass[journal]{IEEEtran}
\usepackage{amsmath,amssymb,amsfonts}
\usepackage{algorithmicx}
\usepackage{algorithm, algpseudocode}
\usepackage{graphicx}
\usepackage{textcomp}
\usepackage{xcolor}
\usepackage{multirow}
\usepackage{amsthm}
\usepackage{amsmath}
\usepackage{mathtools}
\usepackage{cancel}
\usepackage{subcaption}
\usepackage{lipsum}
\usepackage{mathtools}
\usepackage{cuted}
\usepackage{dblfloatfix}
\usepackage{footnote}
\usepackage[autostyle]{csquotes}
\begin{document}
\title{A Blockchain based Federated Learning for Message Dissemination in Vehicular Networks}

\author{\thanks{This project has received funding from the European Union’s Horizon 2020 research and innovation programme under the Marie Skłodowska-Curie grant agreement No 101006411.}
	Ferheen~Ayaz,~\IEEEmembership{Graduate Student Member,~IEEE,}
	Zhengguo~Sheng,~\IEEEmembership{Senior Member,~IEEE,}
	Daxin~Tian,~\IEEEmembership{Senior Member,~IEEE}
	and~Yong~Liang~Guan,~\IEEEmembership{Senior Member,~IEEE}% <-this % stops a space
	\thanks{F. Ayaz and Z. Sheng are with the Department of Engineering and Design, University of
		Sussex, Brighton, BN1 9RH, U.K. e-mail: (f.ayaz@sussex.ac.uk, z.sheng@sussex.ac.uk).}% <-this % stops a space
	\thanks{D. Tian is with School of Transportation Science and Engineering,
		Beihang University, Beijing, 100191, China. e-mail: (dtian@buaa.edu.cn)}
	\thanks{Y. L. Guan is with the School of Electrical and Electronic Engineering, 
		Nanyang Technological University, 639798, Singapore. e-mail: (eylguan@ntu.edu.sg)}}% <-this % stops a space
\maketitle

\begin{abstract}
Message exchange among vehicles plays an important role in ensuring road safety. Emergency message dissemination is usually carried out by broadcasting. However, high vehicle density and mobility usually lead to challenges in message dissemination such as broadcasting storm and low probability of packet reception. This paper proposes a federated learning based blockchain-assisted message dissemination solution. Similar to the incentive-based Proof-of-Work consensus in blockchain, vehicles compete to become a relay node (miner) by processing the proposed Proof-of-Federated-Learning (PoFL) consensus which is embedded in the smart contract of blockchain. Both theoretical and practical analysis of the proposed solution are provided. Specifically, the proposed blockchain based federated learning results in more number of vehicles uploading their models in a given time, which can potentially lead to a more accurate model in less time as compared to the same solution without using blockchain. It also outperforms the other blockchain approaches for message dissemination by reducing 65.2\% of time delay in consensus, improving at least 8.2\% message delivery rate and preserving privacy of neighbor vehicle more efficiently. The economic model to incentivize vehicles participating in federated learning and message dissemination is further analyzed using Stackelberg game model.  

\end{abstract}


% \fi
\IEEEpeerreviewmaketitle
\section{Introduction}
\IEEEPARstart{T}{raditional} Vehicular Ad-hoc NETwork (VANET) is growing into Internet-of-Vehicles (IoV) to manage large amounts of data transmission, computation and storage and to meet the increasing requirements of infotainment and road safety \cite{IoV}. An IoV enables Vehicle-to-Everything (V2X) communications including Vehicle-to-Vehicle (V2V) and Vehicle-to-Infrastructure (V2I) communications. V2I communication usually refers to an infrastructure dependent VANET, where a cellular base station or a Road Side Unit (RSU) is used to provide a real-time and reliable traffic information. However, a large number of RSUs to provide full coverage in urban areas and high traffic densities require huge installation and maintenance cost \cite{RSU}. Therefore, effective and reliable infrastructure-less V2V communication is necessary in emergency situations such as accidents and traffic jams, so that traffic information can be exchanged in real time, even if RSU is out of reach. In V2V communications, multi-hop relaying is one of the challenges to successfully deliver a message over a wide area. Optimal relay selection mechanisms result in better coverage, more reliable connectivity and less communication overhead \cite{relay}. Various intelligent relay selection schemes depending on a vehicle's distance from predecessor, moving direction, speed and propagation loss in environment have been proposed using fuzzy logic \cite{fuzzy} or machine learning algorithms \cite{machine}. Existing literature shows improved packet delivery ratio by machine learning algorithms in multi-hop V2V communications \cite{MARS}. However, artificial intelligence methods require huge processing power and are often not suitable for a fully distributed architecture \cite{Deep}.

In a traditional centralized architecture of machine learning, the data collected by mobile devices is uploaded and processed in a cloud based server to produce inference models \cite{cloud}. With potentially large number of autonomous vehicles, where real-time decisions have to be made within a restricted time period, a cloud-centric approach is unable to offer acceptable latency and scalability. Also, a centralized architecture requires full connectivity which is challenging for vehicular networks. Federated learning (FL) is a distributed machine learning approach, in which mobile devices collect data and train their individual machine learning or deep learning models, called local models. They send their local models (i.e., models' weights) to an aggregator. The aggregator averages local models and produces a global model. Mobile devices further train the global model individually to create updated local models and submit them to aggregator. The steps are repeated in multiple iterations until a desired accuracy of global model is achieved \cite{Google}. FL is considered as a feasible solution for safety and time critical applications involving autonomous vehicles \cite{FL1}. 

Despite offering a distributed approach, FL still relies on a central aggregator. Furthermore, it needs a sustainable economic model to incentivize mobile devices based on their contributions and prevent adversary attacks. For example, in IoV, a malicious vehicle may deliberately modify data, causing poisoning attack \cite{FL2} or a selfish vehicle may not cooperate in data collection resulting in inaccurate weights of a local model. Blockchain can be used with FL to provide a decentralized solution, for managing incentives and ensuring security and privacy in a trustworthy manner \cite{FLBn2}. A blockchain is a distributed ledger of immutable blocks which are added after undergoing a set of rules called consensus \cite{Nakamoto}. Due to its decentralized nature, blockchain complements both FL and IoV \cite{FLB}. Furthermore, smart contracts, which are self executing scripts stored in blockchain to enforce a set of rules, allow automation of multi-step processes and interaction among mobile devices \cite{Smart}. Therefore, they can be used to set rules for protecting FL from adversary and security attacks. The process of transaction verification in blockchain can also be utilized to validate local models in FL \cite{FLBn}. Table \ref{table1} summarizes the current issues of FL in IoV and corresponding solutions provided by blockchain.

Practical implementation of blockchain in vehicular networks is challenging. Due to limited connectivity duration in V2V communications, moving vehicles may not always have an updated blockchain ledger, which leads to possibility of multiple blocks added in parallel, called forks, as shown in Fig.~\ref{fork} (a). With presence of forks, it is difficult for all vehicles in a network to attain a synchronized linear structure of ledger. As a common practice, blockchain picks one of the parallel blocks to continue, and meanwhile, disqualifies other forking blocks by longest chain acceptance protocol \cite{chain}. Forks also lead to creation of malicious chains \cite{prism}. To address this issue, the hierarchical structure of blockchain is proposed for vehicles \cite{heir, PoQF}. In a hierarchical structure, there are two types of blocks: keyblock and microblock. Instead of a linear ledger, microblocks representing off-chain transactions are added in parallel, whereas keyblocks are main blocks which are appended horizontally in a blockchain by a leader or a central node, for example, RSU. As shown in Fig.~\ref{fork} (b), parallel addition of microblocks does not disturb the main linear ledger and forks are not disqualified but accepted as off-chain micro-transactions recorded in a decentralized manner. 

\begin{figure*}
	\begin{subfigure}{0.4\textwidth}
	\includegraphics[scale=0.25]{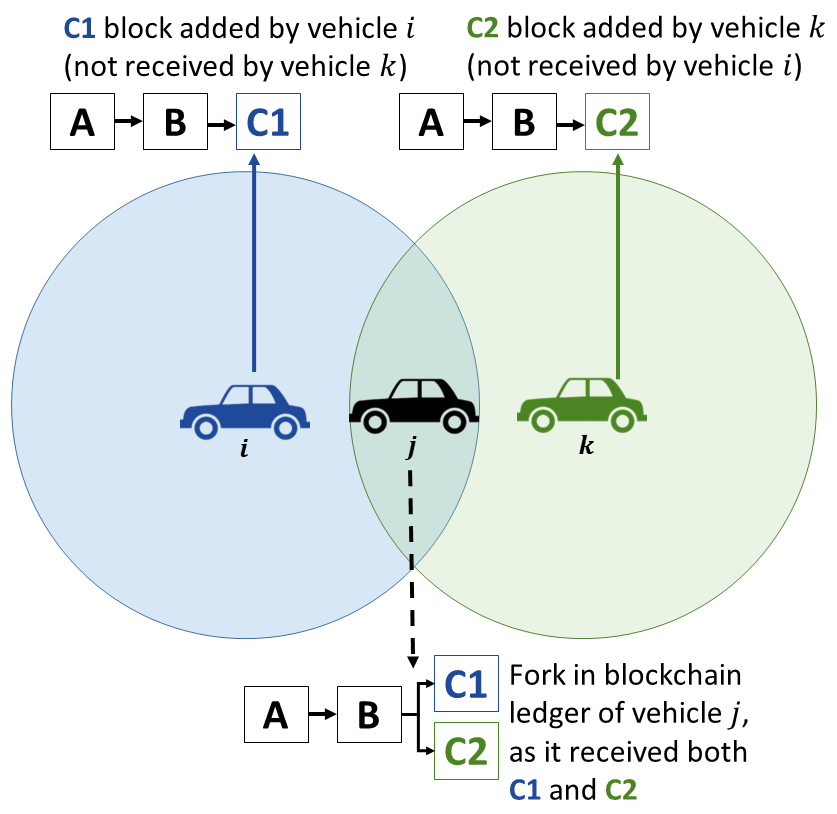}
	\caption{Fork (parallel blocks) in linear blockchain.}
	\end{subfigure}
\hspace*{\fill} % separation between the subfigures
\begin{subfigure}{0.4\textwidth}
	\centering
	\includegraphics[scale=0.25]{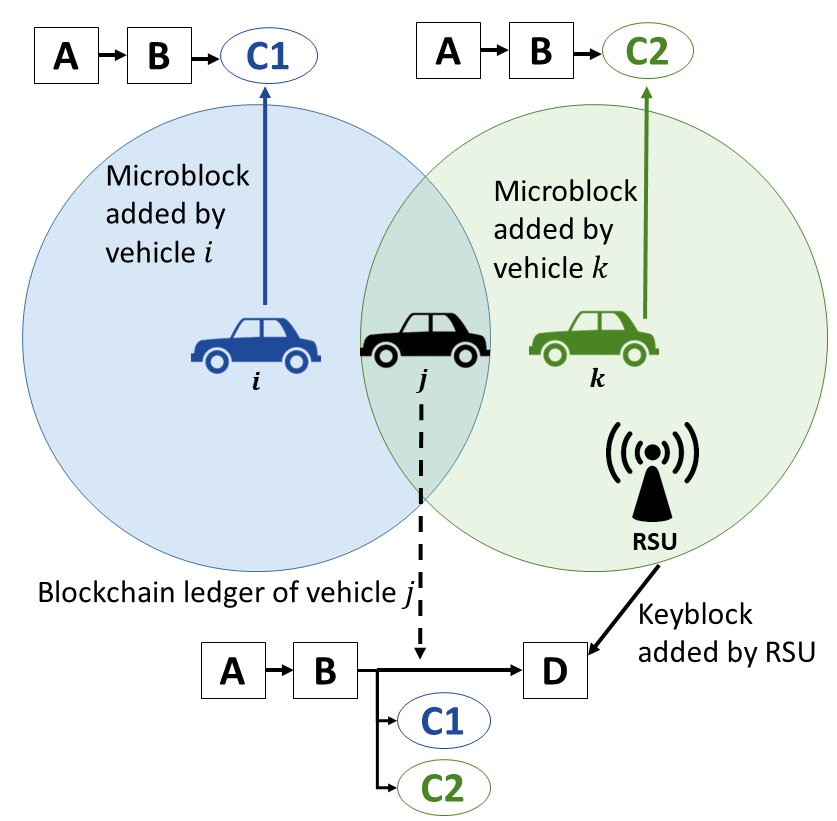}
	\caption{Parallel microblocks and linear keyblock.}
\end{subfigure}
\caption{Parallel addition of microblocks to resolve forks in blockchain in vehicular networks.}
\end{figure*} \label{fork}

\begin{table}
	\renewcommand{\arraystretch}{1.3}
	\caption{FL issues \& blockchain solutions.}
	\label{table1}
	\centering
	\begin{tabular}{|c|c|}
		\hline
		\textbf{FL issue} & \multirow{2}{*}{\textbf{Blockchain-based solution}} \\
		\textbf{in vehicular networks} & {}\\
		\hline
		Requires central aggregator & Independent of third party \\
		Lacks economic modeling & Manages cryptocurrency based incentives \\
		Requires adversary control & Provides security by smart contract \\
		\hline
	\end{tabular}
\end{table}
In this paper, we propose a decentralized message dissemination solution using a hierarchical blockchain based FL process. The vehicles train local models and RSU acts as aggregator to consolidate the global model. The process uses blockchain for updating local models in a decentralized manner. The main contributions of the paper are:
\begin{itemize}
\item We propose a blockchain based FL process in vehicular networks, where a smart contract is used to control adversary attacks by malicious and selfish vehicles. Lower Mean Squared Error (MSE) in less number of iterations is achieved by the global model produced through FL if security check is enforced by the smart contract.  An economic model to incentivize relay nodes and vehicles participating in FL process is also presented together with its analysis using Stackelberg game.
\item Theoretical and simulation analysis in presence and absence of blockchain are presented. In a given time slot, the number of local models uploaded through blockchain based FL are higher than a centralized approach without blockchain, which concludes that the proposed solution can achieve greater accuracy within less time.
\item We propose a Proof-of-FL (PoFL) consensus for a blockchain-based multi-hop relay selection scheme in V2V communications. PoFL results averagely in a reduced time delay per hop by 65.2\%, an improved message delivery ratio by at least 8.2\% and a more privacy-preserving approach compared to other blockchain approaches suitable for message dissemination.
\end{itemize}

The rest of the paper is organized as follows. Section II describes related work. Section III explains the proposed solution of blockchain based FL and PoFL based message dissemination with discussion on its privacy feature. Section IV theoretically analyzes training capacity of FL and the proposed economic model. Simulation results and conclusion are presented in Section V and Section VI respectively.
\section{Related Work}
\subsection{Multi-Hop Relay Selection}
Relay selection in V2V communications plays a crucial role in message broadcasting. An inappropriate relay may cause unacceptable latency or sometimes failure in delivering a message to a desired number of vehicles or area. Probabilistic calculations are usually used to predict either the distance \cite{dist} or link stability \cite{link} of a relay node. In \cite{PoQF}, a Proof-of-Quality-Factor (PoQF) is established using probabilistic estimation of both distance and Signal to Noise Interference Ratio (SINR) as a merit of relay node selection. However, probabilistic predictions rely on certain approximations, for example, number of vehicles within a transmission range, which may not be highly accurate with varying speeds. In \cite{ICC} and \cite{book}, the combination of distance of a vehicle from previous sender and channel quality parameters are used to determine link stability for relay node selection. It is crucial to set weights of all parameters according to their impacts on message delivery in a network. 

To make relay selection more adaptive to network changes, artificial intelligence based mechanisms are designed. Fuzzy logic has been used in \cite{fuzzy} and \cite{fuzzy1}, which makes decision according to distance, moving direction and speed of vehicles. However, fuzzy logic is also dependent on thresholds and weights to be set in the rule base for making inferences. In \cite{machine}, satellite images are used to detect buildings and obstacles to enable machine learning driven channel characterization. The path with lowest propagation loss is used for message dissemination in \cite{machine}. RSU assisted deep learning based technique is developed for relay selection in \cite{Deep}. It is pointed out that machine learning and deep learning algorithms require large processing power to handle huge amount of data and therefore they must require V2I communications and infrastructure support for implementation.
\subsection{FL in Vehicular Networks}
FL is suggested as a promising technique to securely train intelligent models across smart cars \cite{FL1} and Unmanned Aerial Vehicles (UAVs) \cite{FLV1}. It has the feature of reducing network latency by dividing training task among network edges. In cellular-V2X (C-V2X) communications, FL is proposed to reduce failure probability by intelligently offloading high computation tasks to nearby base stations \cite{FLVC1}. Resource allocation and sharing in C-V2X by FL among vehicles has promised better coverage and Quality-of-Service (QoS) in \cite{FLVC2b}. FL and fog-assisted V2X is presented in \cite{FLVC3} to improve driving experience of autonomous vehicles by providing user-end services, for example, car sharing, intelligent parking allocation, infotainment and e-commerce applications. In \cite{FLVC4}, FL is used to tackle energy transfer issues of electric vehicles at charging stations and has resulted in improved accuracy of energy demand prediction. FL assisted blockchain is proposed in \cite{FLBC} to adjust block arrival rate in order to reduce communication latency and consensus delay among vehicles. Applications of FL in vehicular networks are summarized in \cite{FLVC5} and most of the recent applications focus on resource management, performance optimization in computing tasks and user-end services. However, FL can also be promising in message delivery and relay node selection. Table \ref{table2} summarizes the challenges of existing multi-hop relay selection schemes and solutions offered by blockchain-based FL.

\begin{table}
	\renewcommand{\arraystretch}{1.3}
	\caption{Multi-hop relay selection challenges and solutions offered by blockchain-based FL.}
	\label{table2}
	\centering
	\begin{tabular}{|c|c|c|}
		\hline
		\textbf{Approach} & \textbf{Challenge} & \textbf{Solution}\\
		\hline
		Probabilistic & Assumptions / rules  & Local models trained \\  
		prediction \cite{PoQF, link} & are not adaptable  & with different networks \\
		\cline{1-1}
		Fuzzy & to network & and the global model \\
		logic \cite{fuzzy, fuzzy1} & changes & can cater network changes \\
		\hline
		Machine & Huge data have to & Distributed learning and\\
		learning \cite{machine, Deep} & managed centrally &decentralized storage \\
		\hline
		Any scheme without & Relay nodes may & Blockchain incentives \\
		incentives \cite{dist} & act selfish &  for motivation \\
		\hline
	\end{tabular}
\end{table}

\subsection{Economic Modeling in FL}
Economic models to strategize incentives and to promote mobile devices for producing reliable local models have been developed. For a blockchain based FL in \cite{FLBC}, the authors have suggested to incentivize vehicles for both model training and block mining. A joint price and reputation based economic model is proposed in \cite{FL2} to incentivize devices according to the size of data contributed and prevent poisoning attack. The economic model is analyzed using Contract Theory. In Contract Theory, the contracts are formed between a payer and a service provider (i.e., devices training local models) before initiation of FL process. FL among vehicles for image classification is proposed in \cite{FLimage} and contracts are formulated to incentivize vehicles in proportion to the number of images used and amount of computation resources consumed.  

Stackelberg game approach is used in \cite{Stalk1} and \cite{Stalk2} to analyze the actions of players when incentives are distributed after FL iterations are completed. If relay nodes are involved in incentive distribution among vehicles, the economic model is more suited to be analyzed using Stackelberg game model. Due to varying speed and position of vehicles, it is practically better to select an appropriate relay node after a message is originated. Therefore, analysis using Stackelberg game model is a more feasible option for multi-hop relay selection scheme than Contract Theory, because formation of contract prior to FL process initiation or relay selection cannot be materialized.  The existing literature assumes information asymmetry, i.e. the payer is not aware of the amount of contributions (for example, data size) upon which the payment is to be made. However, with public blockchain, where stored transactions are visible to every member of blockchain, FL can be a case of symmetric information, i.e., the relevant information is known to all associated members.  
\begin{table}
	\renewcommand{\arraystretch}{1.3}
	\caption{Key notations.}
	\label{table3}
	\centering
	\begin{tabular}{|c|c|}
		\hline
		\textbf{Notation} & \textbf{Definition} \\
		\hline
		$ORG$ & Originator vehicle\\
		$RLY$ & Relay node vehicle \\
		$k$ & Iteration index \\
		$\pmb{w}_x^k $ & Weights of model $x$ (local or global) at $k^{th}$ iteration \\
		$L(\pmb{w}_x^k)$ & Loss function of model $x$ at $k^{th}$ iteration \\
		$TS$ & Time slot to upload local models \\
		$R$ & Transmission range\\
		$\lambda_{MB}$ & Microblock arrival rate per second \\
		$\lambda_V$ & Vehicle density per m$^2$ \\
		$\mu_d$ & Average distance of vehicles from RSU \\
		$\mu_v$ & Average speed of vehicles \\
		$E(.)$ & Expected value \\
		$N$ & No. of vehicles participating in FL\\
		$N_B$ & No. of vehicles uploading local models via FL blockchain \\
		$N_{WB}$ & No. of vehicles uploading local models without blockchain \\
		$N_{V}$ & No. of vehicles with RSU in transmission range \\
		$N_{MV}$ & No. of moving vehicles reaching RSU \\
		$N_{RLY}$ & No. of relay nodes \\
		$s_i$ & Data size of vehicle $i$ \\
		$I$ & Incentive  \\
		$C(s_i)$ & Cost of training on data of size $s_i$ \\
		$U_x$ & Utility of $x$ (vehicle $i$ or $RLY$) \\
		$\alpha_i$ & Cost coefficient of vehicle $i$ \\
		$\beta$ & Compensation  \\
		$p_m$ & Probability of using $s_m$ \\
		$M$ & Number of possible $s_m$ \\
		\hline
		\multicolumn{2}{|c|}{\textbf{Dataset}}\\
		\hline
		$d_{i,x}$ & Distance between vehicle $i$ and $x$ (vehicle or RSU) \\ 
		$dir_{i,s}$ & Direction of vehicle $i$ w.r.t sender $s$ \\
		$v_i$ & Speed of vehicle $i$ \\
		$h$ & Hop index \\
		$\gamma_i$ & Traffic density in transmission range of vehicle $i$ \\
		$N_A$ & No. of acknowledgment messages \\
		\hline
	\end{tabular}
\end{table}

\section{The Proposed Solution}
As shown in Fig.~\ref{figblockd}, the proposed solution consists of two major parts: 
\\
(1) FL integrated with blockchain, where vehicles take part in a blockchain based FL process to form a global model for relay node selection, and
\\ 
(2) PoFL based message dissemination, where the global model produced in first part is run by vehicles as PoFL to find their eligibility to become a relay node. Table \ref{table3} lists the key notations used in this paper.
\subsection{Blockchain based FL}
\paragraph{FL Components}
\begin{itemize}
\item \textbf {\textit{Hello} Packet by designated vehicle: }
A Central Authority appoints some designated vehicles to regularly originate a \textit{Hello} packet and share their position to initiate dataset collection by vehicles participating in FL. Only the designated vehicles are allowed to originate \textit{Hello} packets. The motivation behind designated vehicles is two fold: first is because they are trusted by Central Authority to honestly send their actual position without any malicious change and second is because the identities of designated vehicles are already shared with other vehicles, so \textit{Hello} packet from any other identity is not recognized by the network. 
\begin{figure}
	\centering
	\includegraphics[scale=0.18]{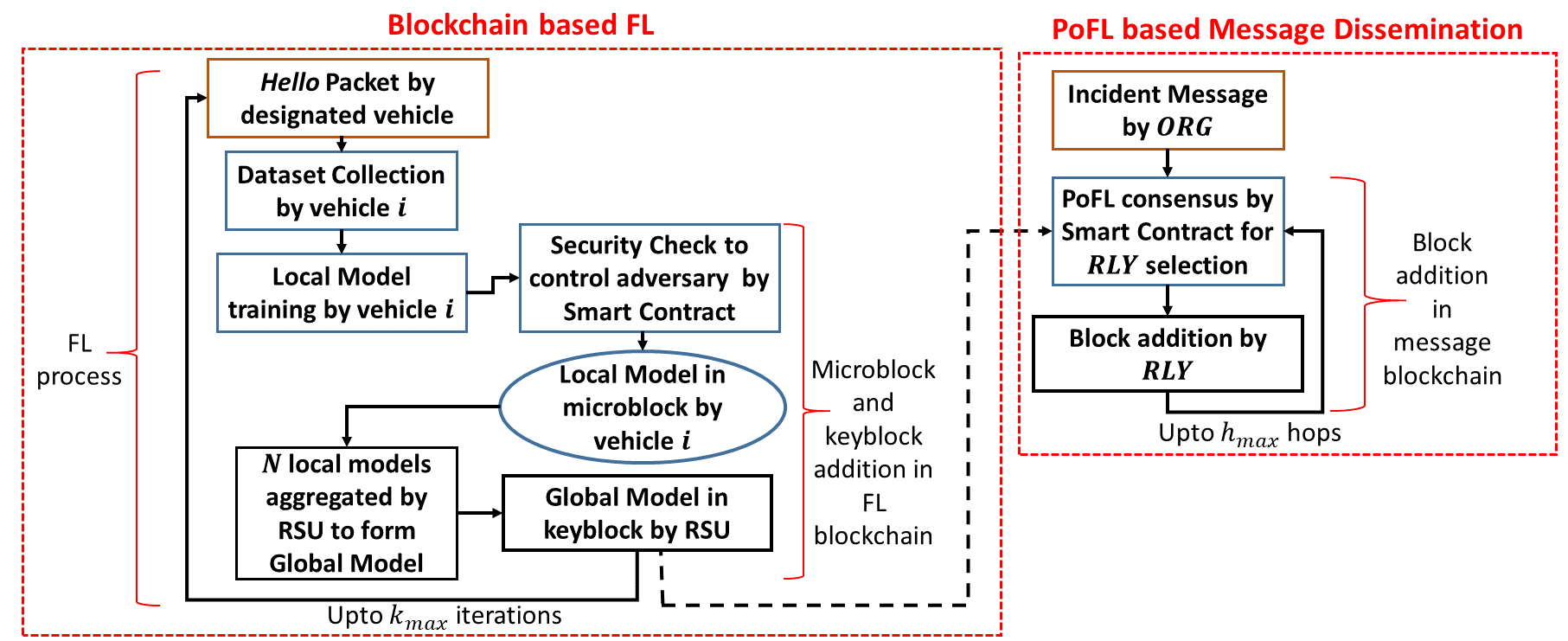}
	\caption{The proposed steps by vehicles and RSU.}
	\label{figblockd}
\end{figure}
\begin{figure*}
	\centering
	\includegraphics[scale=0.25]{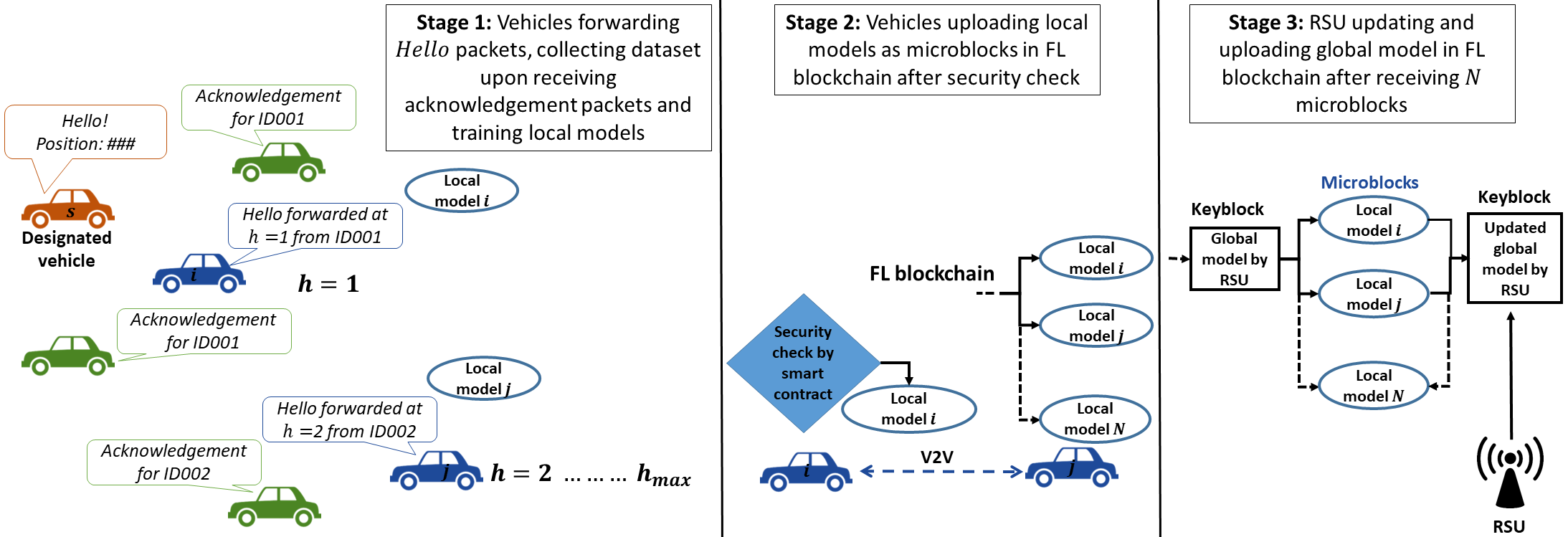}
	\caption{The proposed stages in blockchain based FL.}
	\label{fig1}
\end{figure*}
\item \textbf{Dataset: } 
It refers to the data samples collected by vehicle $i$ for training local model. In the proposed solution, dataset collected by vehicle $i$ includes multi-hop relay selection parameters. After forwarding a \textit{Hello} packet, dataset collected by vehicle $i$ consists of the following parameters mentioned in Table~\ref{table3}: $d_{i,s}$, distance from sender $s$ (designated vehicle or previous relay node). $dir_{i,s}$, moving direction (either towards or away from sender $s$). $v_i$, speed at the time of forwarding message. $h$, hop number. $\gamma_i$, traffic density within its transmission range and $N_A$, number of acknowledgments received as the \textit{score} of relaying. $\gamma_i$ in dataset can be pre-specified by Central Authority or estimated by counting average number of vehicles sending beacon messages per meter \cite{beacon}. The process of dataset collection is explained in detail later in this section.
\item \textbf{Local Model: }
Each vehicle $i$ participating in FL trains a Deep Neural Network based local model.
\item \textbf{Global Model: }
It is an aggregated model of $N$ local models, where $N$ also refers to the number of vehicles participating in FL. In our proposed system, the global model is consolidated by RSU.
\end{itemize}
\paragraph{FL Blockchain and its Components}
FL blockchain is a blockchain used by vehicles and RSUs to store local and global models as blocks. Its main components include
\begin{algorithm} [t]
	\caption{FL Algorithm for vehicle $i$}\label{alg1}
	\algnewcommand{\algorithmicgoto}{{Go to}}%
	\algnewcommand{\Goto}[1]{\algorithmicgoto~\ref{#1}}%
	\hspace*{\algorithmicindent}
	\textbf{Input:} \textit{Hello} Packet, $N$ vehicles \\ \hspace*{\algorithmicindent}
	\textbf{Output:} Global Model
	\begin{algorithmic}[1]
		\While {$h \leq h_{max}$}  \label{hello}
		\State Generate random waiting time
		\While {Time elapsed $\leq$ random waiting time}
		\If {Forwarded \textit{Hello} packet received at $h$}
		\State break
		\EndIf
		\EndWhile
		\If {Forwarded \textit{Hello} packet not received at $h$}
		\State Forward \textit{Hello} packet
		\State Count acknowledgment packages into $N_A$ 
		\State Record $d_{i,s}$, $v_i$, $dir_{i,s}$, $\gamma_i$, $N_A$ in dataset
		\State break
		\Else
		\State $h=h+1$
		
		\EndIf
		\EndWhile
		\If {data size $==s_i$}
		\State Train local model
		\Else
		\State \Goto{hello}
		\EndIf
		\While{$k \leq k_{max}$}
		\State Upload local model through smart contract
		\State Receive updated global model
		\State Re-train local model
		\State $k=k+1$
		\EndWhile
	\end{algorithmic}
\end{algorithm}
\begin{itemize}
\item \textbf{Adversary: }
We consider the following adversary threats:
\begin{itemize}
	\item \textit{Malicious Vehicles:} They may deliberately change or inject false data so that local model is not trained accurately. This phenomenon is also known as poisoning attack \cite{FL2}. 
	\item \textit{Selfish Vehicles:} They may not send acknowledgment messages despite receiving forwarded messages. Therefore, $N_A$ cannot be recorded correctly during dataset collection, leading to an inaccurate local model produced by vehicle $i$. 
\end{itemize}
\item \textbf{Security Check: }
It is a machine learning algorithm embedded in smart contract of FL blockchain to detect adversary before a local model is uploaded as a block by vehicle $i$.
\item \textbf{Microblock: }
A local model is stored in FL blockchain as a microblock after undergoing a security check. A microblock is added in parallel to other microblocks, all containing hash of previous keyblock. 
\item \textbf{Keyblock: }
A global model is stored in FL blockchain by RSU in the form of a keyblock, containing hashes of previous $N$ microblocks. 
\end{itemize}
As shown in Fig.~\ref{fig1}, the proposed blockchain based FL consists of the following three stages:
\subsubsection{Stage 1: Dataset Collection and Local Model Training}
In this stage, vehicles collect dataset for training. Upon receiving a \textit{Hello} packet from a designated vehicle, a vehicle $i$ which aims to collect dataset, generates a random waiting time. When the waiting time is complete, it forwards \textit{Hello} packet with its encrypted identity. The reason behind a random waiting time is to prevent multiple vehicles from transmitting at the same time and avoid packet collision. The limits and probability distribution of random waiting time are described in \cite{PoQF}. The vehicles which receive the forwarded \textit{Hello} packet for the first time share their acknowledgment. An acknowledgment packet contains encrypted identity of vehicle $i$, so that it can collect dataset. A vehicle $j$, which participates in FL, will broadcast the received \textit{Hello} packet again after a random waiting time. This process continues up to a specified number of hops, $h_{max}$, as shown in Algorithm~\ref{alg1}. Each vehicle produces a local model based on Deep Neural Network with 7 hidden layers and 256 neurons in each layer. 
\subsubsection{Stage 2: Security Check and FL Blockchain Update}
A vehicle $i$ shares its local model with the network by adding it into FL blockchain as a microblock. It is added after passing a security check performed by the smart contract embedded in FL blockchain. The proposed security check employs a machine learning algorithm called Isolation Forest \cite{isolationforest} to detect anomaly in a local model caused by adversary. Isolation Forest is used because it only requires a small number of samples for training. A true sample is provided by the Central Authority for its initial training. Later, it can be used in a fully unsupervised manner to detect anomaly. Moreover, it is computationally efficient and has low memory requirement \cite{isolation2}. We have used the security check in three ways. Firstly, the security check on dataset is conducted by finding anomalies in dataset of each vehicle. Secondly, the security check performs anomaly detection on weights of local models. If a malicious vehicle $i$ deliberately changes its dataset for training its local model but shares a true dataset in smart contract, the adversary attack will be detected by anomaly detection on weights. Thirdly, the security check on both dataset and weights is performed. If local models successfully pass the security check, they are added in FL blockchain in the form of parallel microblocks. The microblock announcement is broadcasted by vehicle $i$ and the receiving vehicles will then update their copy of FL blockchain. Vehicles can exchange new microblock updates with their neighbors regularly.
\subsubsection{Stage 3: Global Model Aggregation by RSU}
Whenever a vehicle $i$ finds an RSU available in its transmission range, it shares its updated copy of FL blockchain. When $N$ microblocks are received by RSU in FL blockchain, it aggregates local models into a global model and uploads it into a keyblock. 

All stages are repeated at each iteration. The goal is to repeat the process up to $k_{max}$ iterations for minimizing global loss function $L(\pmb{w}_G^k)$, which is defined as
\begin{equation}
L(\pmb{w}_G^k) = \frac{1}{N} \sum_{i=1}^N  L(\pmb{w}_i^k).
\end{equation} 
where $\pmb{w}_G^k$ are weights of global model and $L(\pmb{w}_i^k)$ is the loss function of local model $i$ and $\pmb{w}_i^k$ are its corresponding weights at $k^{th}$ iteration. Neural networks commonly use MSE as the loss function \cite{FL1}. The value of $k_{max}$ is adjusted by Central Authority to achieve the minimum possible or desired $L(\pmb{w}_G^{k_{max}})$ \cite{Google}.
\begin{algorithm}[t]
	\caption{Message Dissemination Algorithm for vehicle $i$}\label{alg2}
	\algnewcommand{\algorithmicgoto}{{Go to}}%
	\algnewcommand{\Goto}[1]{\algorithmicgoto~\ref{#1}}%
	\hspace*{\algorithmicindent}
	\textbf{Input:} Incident message, global model \\ \hspace*{\algorithmicindent}
	\textbf{Output:} New block announcement in message blockchain
	\begin{algorithmic}[1]
		\While {$h \leq h_{max}$}
		\State Compute \textit{score} from global model 
		\State timer expiry limit = $1/score$
		\While {Time elapsed $\leq$ timer expiry limit}
		\If {New block announced}
		\State break
		\EndIf
		\EndWhile
		\If {New block not announced}
		\State Announce block
		\State break
		\Else
		\State $h=h+1$
		\EndIf
		\EndWhile
	\end{algorithmic}
\end{algorithm}
\subsection{PoFL based Message Dissemination}
The main components of this part include
\begin{itemize}
\item \textbf{Incident Message: }It is a message initiated by an originating vehicle ($ORG$) in an emergency situation, for example, incident or traffic jam. It contains time and position of incident and encrypted identity of $ORG$.
\item \textbf{PoFL: }It is the consensus to elect a relay node ($RLY$) for forwarding an incident message. Global model contained in the latest keyblock of FL blockchain is used as PoFL. It is run by a smart contract of message blockchain.
\item \textbf{Message Blockchain: } It contains history of incident messages. When a $RLY$ is elected by PoFL, it adds a block in the message blockchain containing the forwarded incident message. The block also contains time, location and encrypted identity of the $RLY$ which adds the block. The motivation behind message blockchain is to record forwarded incident messages as immutable blocks and avoid discrepancies during incentive distribution among $RLYs$. 
\end{itemize}
\begin{figure}
	\centering
	\includegraphics[scale=0.25]{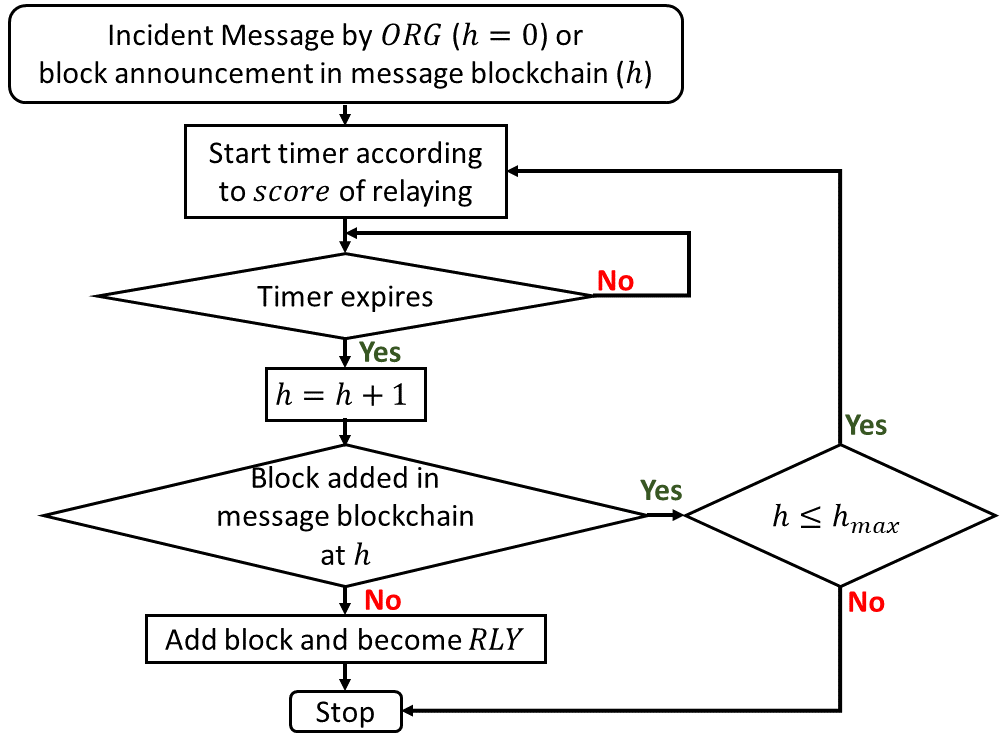}
	\caption{Flowchart of actions by vehicle $i$ according to PoFL based Message Dissemination.}
	\label{msg2}
\end{figure}
When an incident message is initiated by $ORG$, all receiving vehicles attempt to become the $RLY$ by competing through PoFL consensus. Each vehicle $i$ runs PoFL consensus to find its \textit{score} of being $RLY$, as shown in Algorithm~\ref{alg2}. PoFL is aimed to assign the highest \textit{score} to the most appropriate relay node. The smart contract starts a timer whose length is inversely proportional to the \textit{score} of vehicle $i$. As shown in Fig.~\ref{msg2}, a block is added in the message blockchain and a block announcement with the forwarded incident message is initiated by vehicle $i$ if its timer first expires. In this case, vehicle $i$ is assigned as a relay node $RLY$ at $h=1$. All other vehicles continue to compete for becoming $RLY$ at further hops until the message is forwarded up to $h_{max}$ number of hops. 
\paragraph{Privacy of PoFL based Message Dissemination}
\begin{table}
	\renewcommand{\arraystretch}{1.3}
	\caption{Parameter-sharing required from neighbor vehicles in multi-hop relay selection.}
	\label{table4}
	\centering
	\begin{tabular}{|c|c|c|c|}
		\hline
		\textbf{Approach}  & \textbf{Position} & \textbf{Speed} & \textbf{Other Parameters}\\
		\hline
		Deep learning \cite{Deep} & $\checkmark$ & $\checkmark$ & Transmission power \\
		Fuzzy logic \cite{fuzzy1} & $\checkmark$ & $ \times $  & $ \times $ \\
		Probabilistic prediction \cite{link} & $\checkmark$ & $\checkmark$ & $ \times $ \\
		Link Stability \cite{ICC} & $\checkmark$  & $ \times $ & Channel quality  \\
		PoQF \cite{PoQF} & $\checkmark$ & $\checkmark$ & $ \times $ \\
		PoFL & $ \times $ & $ \times $ & $ \times $ \\
		\hline
	\end{tabular}
\end{table}
Table~\ref{table4} lists the parameters required to be shared by neighbor vehicles in various multi-hop relay selection approaches. The position, speed and heading direction of vehicles are regularly shared in VANETs using beacon messages and thus create a threat to privacy \cite{priv}. The proposed approach does not require such information from all neighbor vehicles and can therefore be considered as a privacy-preserving solution. The position and direction of only sender is required for dataset collection in blockchain based FL and for calculating \textit{score} of relaying in PoFL based message dissemination. However, identities of vehicles are kept anonymous using encryption. 

\section{Theoretical Analysis}
\subsection{Training Capacity of FL}
This subsection is aimed to analyze the capacity of FL blockchain to complete one FL iteration in a given amount of time, compared with the same process without blockchain. FL without blockchain is referred to as a centralized approach in which each vehicle $i$ submits its local model directly to RSU instead of uploading it into FL blockchain. As the convergence performance of FL improves with increasing number of local models \cite{conv}, FL via blockchain is expected to achieve greater accuracy, provided the number of uploaded local models are higher as compared to the process carried out without blockchain within the same time period.

\subsubsection{FL with blockchain} Let $TS$ be a time slot in which a vehicle $i$ is required to upload its local model as microblock in FL blockchain, after it has completed training and passed its local model through security check. Let $\lambda_{MB}$ be the microblock arrival rate at RSU or throughput in microblocks/s. Detailed derivation of $\lambda_{MB}$ can be found in \cite{PoQF}. If microblock arrival is modeled using Poisson distribution as defined in \cite{prism}, the expected number of vehicles able to upload thier local models via FL blockchain in $TS$ can be given as \cite{IoT}
\begin{equation}
E(N_{B}) = \sum_{l=1}^{\lambda_{MB} TS} l e^{- \lambda_{MB} TS} \frac{(\lambda_{MB}TS)^l}{l!} .
\label{eqB}
\end{equation}

\subsubsection{FL without blockchain} 
\begin{figure}
	\centering
	\includegraphics[scale=0.2]{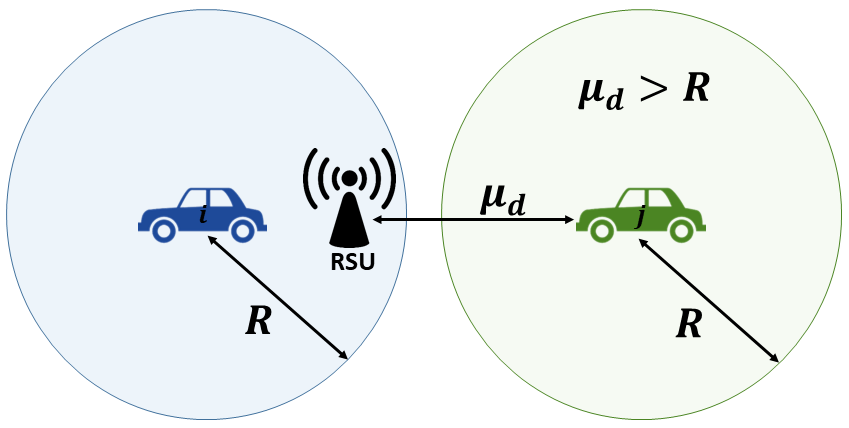}
	\caption{Distance between vehicle and RSU.}
	\label{anal1}
\end{figure}

If vehicles are required to upload their models directly to RSU without blockchain, then it is necessary that either RSU is in their transmission range or they are able to reach towards RSU within $TS$. Consider a general and dynamic movement of vehicles, the position of vehicles on road follows Poisson distribution, $\lambda_{V}$ vehicles/m$^2$ is assumed as the density of vehicles on a two dimensional road segment with no separation of lanes \cite{sKim}, the expected number of vehicles with RSU in their transmission range is
\begin{equation}
E(N_{V}) = \sum_{l=1}^{\lambda_{V} \pi R^2 } l e^{- \lambda_{V} \pi R^2} \frac{(\lambda_{V} \pi R^2)^l}{l!} ,
\label{eqV1}
\end{equation}
where transmission range is assumed as a uniform circle with radius $R$. Similarly, the expected number of moving vehicles with RSU not currently in their transmission range but can travel to reach RSU within $TS$ is
\begin{equation}
E(N_{MV}) = \sum_{l=1}^{\frac{TS}{\mu_d - R} \mu_v } l e^{- \frac{TS}{\mu_d - R} \mu_v} \frac{(\frac{TS}{\mu_d - R} \mu_v)^l}{l!} ,
\label{eqV2}
\end{equation}
where $\mu_d>R$ is the mean distance of those vehicles from RSU which do not have RSU within their transmission range, as shown in Fig.~\ref{anal1} and $\mu_v$ is their average speed. Therefore, the expected number of vehicles able to upload their local models to RSU without blockchain during $TS$ is
\begin{equation}
E(N_{WB})=E(N_{V}) + E(N_{MV}).
\label{eqWB}
\end{equation}
\subsection{Economic Model}
In this section, we define an economic model of payment to vehicles contributing in FL and message dissemination. The feasibility of economic model is analyzed by investigating strategic behavior of $RLYs$ and vehicles participating in FL based on their expected utilities through Stackelberg game model. 
%\begin{figure}
%	\centering
%	\includegraphics[scale=0.25]{stalk.PNG}
%	\caption{The proposed economic model.}
%	\label{stalk}
%\end{figure}
\begin{table}
	\renewcommand{\arraystretch}{1.3}
	\caption{Reward gained and payment made by players.}
	\label{tableStalk}
	\centering
	\begin{tabular}{|c|c|c|}
		\hline
		\textbf{Player}  & \textbf{Gains} & \textbf{Pays} \\
		\hline
		$ORG$ & None & $N_{RLY} \beta log (1+I)$ to $RLYs$\\
		$RLY$ & $\beta log (1+I)$ from $ORG$ & $\sum_{i=1}^N Is_i$ to $N$ vehicles \\
		Vehicle $i$ & $N_{RLY}Is_i$ from $RLYs$ & $\alpha_is_i^2$ to train local model \\
		\hline
		\end{tabular}
\end{table}
\color{black}
\subsubsection{Stackelberg Game Formulation}
The Stackelberg game model consists of three types of players: $ORG$, $RLY$ participating in message dissemination and vehicle $i$ participating in FL. For each incident message initiated by $ORG$, there are $N_{RLY}$ number of $RLYs$ which forward the incident message and $N$ vehicles in the network which train their local models during blockchain based FL. The proposed economic model is formulated as a two-stage Stackelberg game. First, at stage 1, $ORG$ pays reward to $RLYs$ for forwarding message. At stage 2, $RLYs$ pay reward to $N$ vehicles for participating in FL to form a global model of $RLY$ selection. Since, both FL and message dissemination processes are blockchain-based, the contribution of players is stored as immutable timestamped blocks and cannot be altered for cheating. The transactions of incentives are also processed in the form blockchain based virtual currency automatically through smart contracts.

As shown in Table~\ref{tableStalk}, the reward to $N$ vehicles is paid in proportion to the sizes of dataset they have used in training their local models. Assume that the dataset sizes of $N$ vehicles are $\pmb{s}= \{s_1, s_2, ....,s_N \}$. The utility of each vehicle $i$ participating in FL process is
\begin{equation}
U_i (s_i, I) = N_{RLY} I s_i - C(s_i),
\label{eqUi}
\end{equation}
where $I$ denotes incentive which is constant for every vehicle $i$ and $C(s_i)$ is the computational cost of training a local model on dataset of size $s_i$ and is modeled as a quadratic function, i.e.,
\begin{equation}
C (s_i) = \alpha_is_i^2 ,
\end{equation}
where $\alpha_i>0$ denotes cost co-efficient of vehicle $i$ \cite{incentive}. The utility of each $RLY$ is
\begin{equation}
U_{RLY} (\pmb{s}, I) =   \beta log(1+I) - \sum_{i=1}^N I \cdot s_i,
\label{eqURLY}
\end{equation}
where $\beta log(1+I)$ is paid by $ORG$ for forwarding the incident message. Here $\beta>0$ and can be assumed as a compensation amount paid to $RLYs$ present in an area affected by an incident or traffic jam caused by $ORG$. 

\subsubsection{Stackelberg Game Analysis}
We consider the case of information symmetry where every $RLY$ knows data size used by each vehicle $i$ prior to forwarding a message. 
\\
\\
\textbf{Definition 1:} Assume that $s_i^*$ is the optimal data size for each vehicle $i$ and $I^*$ is the optimal incentive amount per data size paid by each $RLY$ to vehicle $i$, then $(s_i^*, I^*)$ is the Nash equilibrium point which satisfies the following conditions
\begin{equation}
U_i(s_i^*, I^*) \geq U_i (s_i, I^*),
\end{equation}
and
\begin{equation}
U_{RLY} (s_i^*, I^*) \geq U_{RLY} (s_i^*, I).
\end{equation}
\\
\\
\textbf{Theorem 1:} There exists a Nash equilibrium point for a vehicle $i$ with $U_i$ defined in (\ref{eqUi}).
\\
\\
\textit{Proof:} For a fixed $I^*$, $U_i$ is 
\begin{equation}
U_i (s_i, I^*) = N_{RLY} \cdot I^* \cdot s_i - \alpha_is_i^2.
\label{eqUi2}
\end{equation}
The first-order derivative of (\ref{eqUi2}) is
\begin{equation}
\frac{\partial U_i (s_i, I^*)}{ \partial s_i} = N_{RLY} \cdot I^* - 2 \alpha_is_i .
\end{equation}
The second-order derivative of (\ref{eqUi2}) is
\begin{equation}
\frac{\partial^2 U_i (s_i, I^*)}{ \partial s_i^2} = -2 \alpha_i .
\end{equation}
Since $\alpha_i>0$, the second-order derivative of $U_i$ is negative and $ U_i (s_i, I^*)$ is a strictly concave function, which proves the existence of Nash equilibrium. \hspace*{\fill} $\qed$
\\
\\
\textbf{Theorem 2:} There exists a Nash equilibrium point for $RLY$ with $U_{RLY}$ defined in (\ref{eqURLY}).
\\
\\
\textit{Proof:} The first-order derivative of (\ref{eqURLY}) is
\begin{equation}
\frac{\partial U_{RLY} (\pmb{s}, I)}{ \partial I} = \frac{\beta}{1+I} - \sum_{i=1}^N s_i .
\end{equation}
The second-order derivative of (\ref{eqURLY}) is
\begin{equation}
\frac{\partial^2 U_{RLY} (\pmb{s}, I)}{ \partial I^2} = \frac{- \beta}{(1+I)^2} .
\end{equation}
Since $\beta>0$ and $(1+I)^2 > 0$, the second-order derivative of $U_{RLY}$ is negative and $U_{RLY} (\pmb{s}, I)$ is a strictly concave function, which proves the existence of Nash equilibrium. \hspace*{\fill} $\qed$
\\
\\
Based on Theorem 1 and Theorem 2, we can state that the unique Stackelberg Nash equilibrium point of our model exists. The Central Authority is responsible to choose values of $I$ and $\beta$ such that Nash equilibrium points for all $N$ vehicles and $RLYs$ become their best response strategies (i.e, $U_i>0$ and $U_{RLY}>0$) and all players are willing to cooperate in the proposed game. 

The proposed economic model assumes information symmetry, i.e., all players have complete information about $s_i$. If private blockchain is used, complete information may not be visible to every player and the economic model will be information asymmetric. In this case, players may predict information through a machine learning method \cite{bayes} or using probabilistic assumption \cite{incentive}. Let $\pmb{s} = \{s_1, s_2, ....s_M \}$ be the sizes of dataset used by vehicles in FL and $p_m$ be the probability that a vehicle $i$ uses $s_m$. (\ref{eqURLY}) can be modified as
\begin{equation}
U_{RLY} (\pmb{s}, I) =   \beta log(1+I) - \sum_{m=1}^M p_mN I s_m,
\label{eqURLY2}
\end{equation}
where $M$ is the total number of possible $s_m$. Similar to Theorem 2, the existence of Nash Equilibrium point can be proved for $U_{RLY}$ defined in (\ref{eqURLY2}). \color{black}

\section{Results and Discussion}
In this section, we discuss simulation results of the proposed solution using OMNeT++, Python and SUMO (Simulation of Urban Mobility). An open-source framework VeINS (Vehicles In Network Simulation) is used to integrate SUMO with OMNeT++ \cite{veins}. Python is employed for carrying out FL using Tensorflow library of machine learning. Python can be embedded into a C++ program by writing an extension module \cite{Python}. Since OMNeT++ is a modular C++ based network simulator, it supports dynamic loading of Python script at run time. The simulation parameters used are listed in Table~\ref{tablesim}.
\begin{table}
	\renewcommand{\arraystretch}{1.3}
	\caption{Simulation Parameters.}
	\label{tablesim}
	\centering
	\begin{tabular}{|c|c|c|c|}
		\hline
		\textbf{Parameters} & \textbf{Values} & \textbf{Parameters} & \textbf{Values}\\
		\hline
		Simulation Time & 300\,s & Protocol & IEEE 802.11p \\
		Size of area & 2.5\,km\,$\times$2.5\,km & Encryption & SHA-256\\
		Data rate & 6\,Mbps & $s_i$ & 8000  \\
		Mobility model & Krauss  &	Loss function & MSE \\
		Number of RSUs & 1 &  $R$ & 250\,m \\ 
		Number of vehicles & 100, 200, 300 & $h_{max}$ & 6 \\
		$k_{max}$ & 100, 110 & $\mu_v$ & 50\,km/hr  \\
		\hline
	\end{tabular}
\end{table}
\begin{figure*}
	\begin{subfigure}{0.45\textwidth}
		\includegraphics*[scale=0.2]{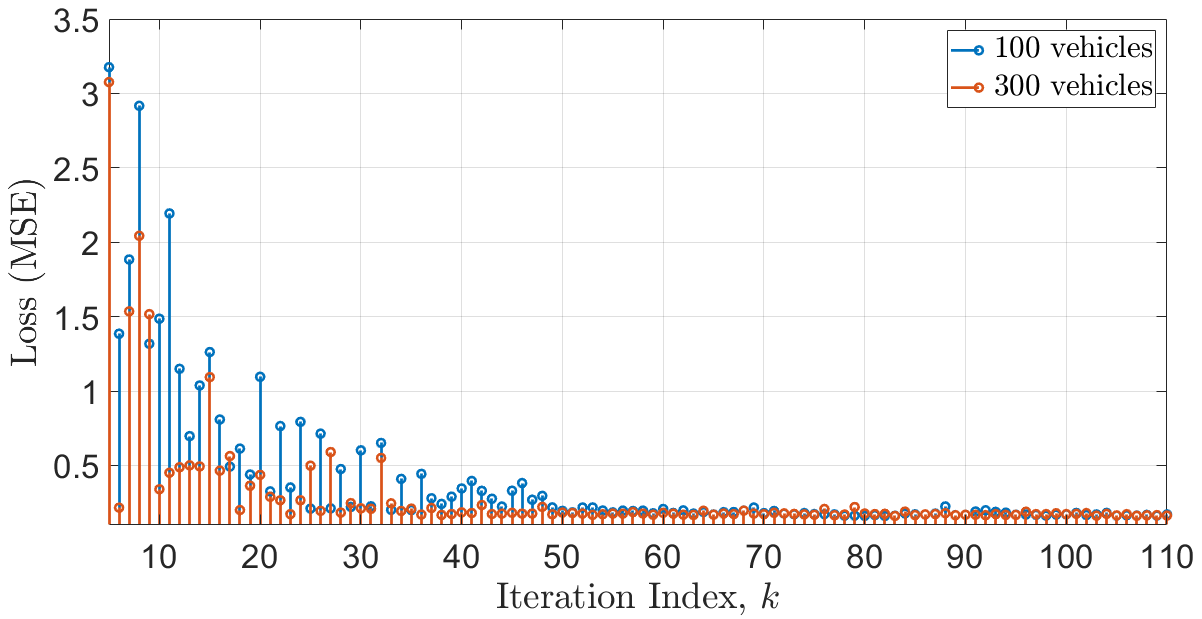}
		\caption{No security Check.}
	\end{subfigure}
	\hspace*{\fill} % separation between the subfigures
	\begin{subfigure}{0.45\textwidth}
		\includegraphics*[scale=0.2]{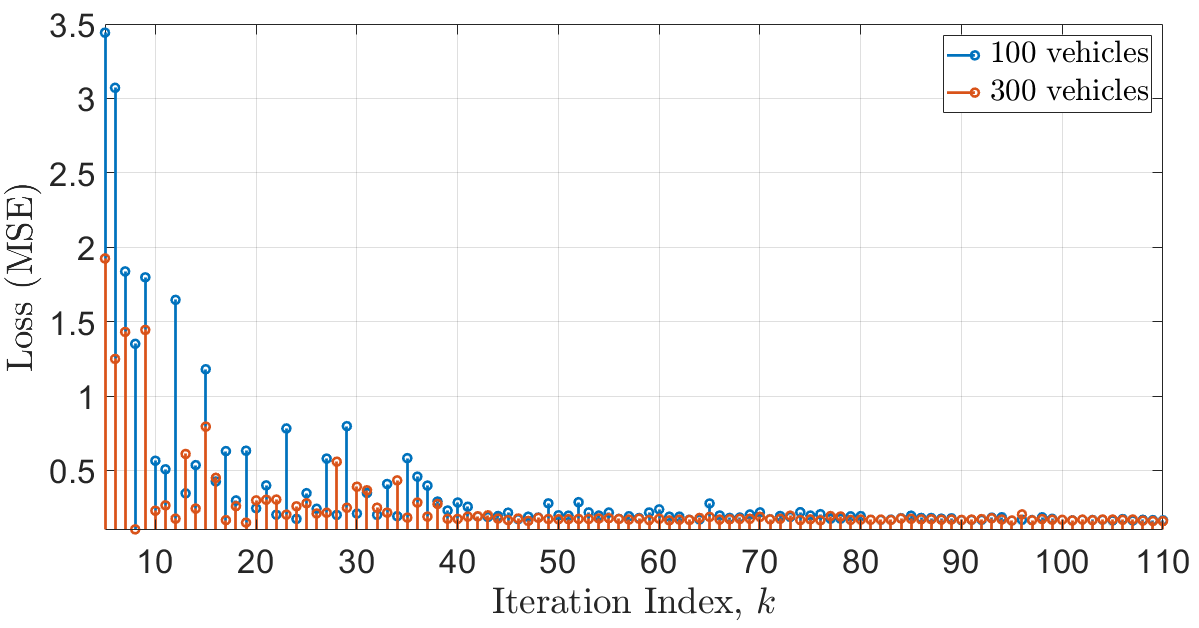}
		\caption{Security Check on dataset.} 
	\end{subfigure}
%	\hspace*{\fill} % separation between the subfigures
	\begin{subfigure}{0.45\textwidth}
		\includegraphics*[scale=0.2]{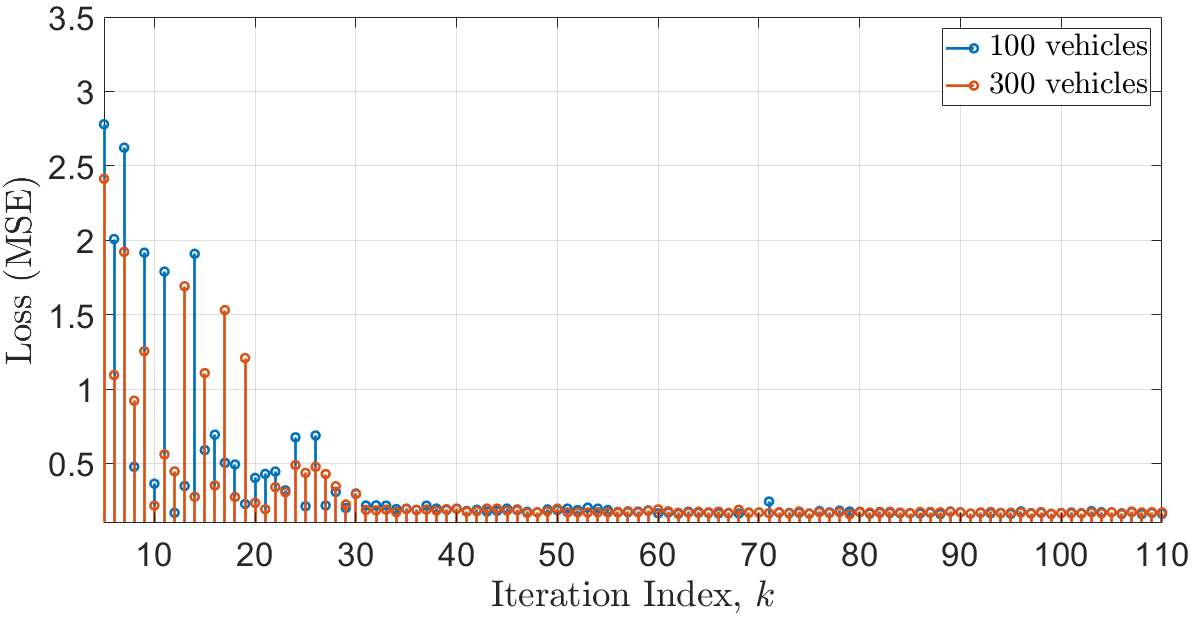}
		\caption{Security check on weights.} 
	\end{subfigure}
	\hspace*{\fill}
	\begin{subfigure}{0.45\textwidth}
		\includegraphics*[scale=0.2]{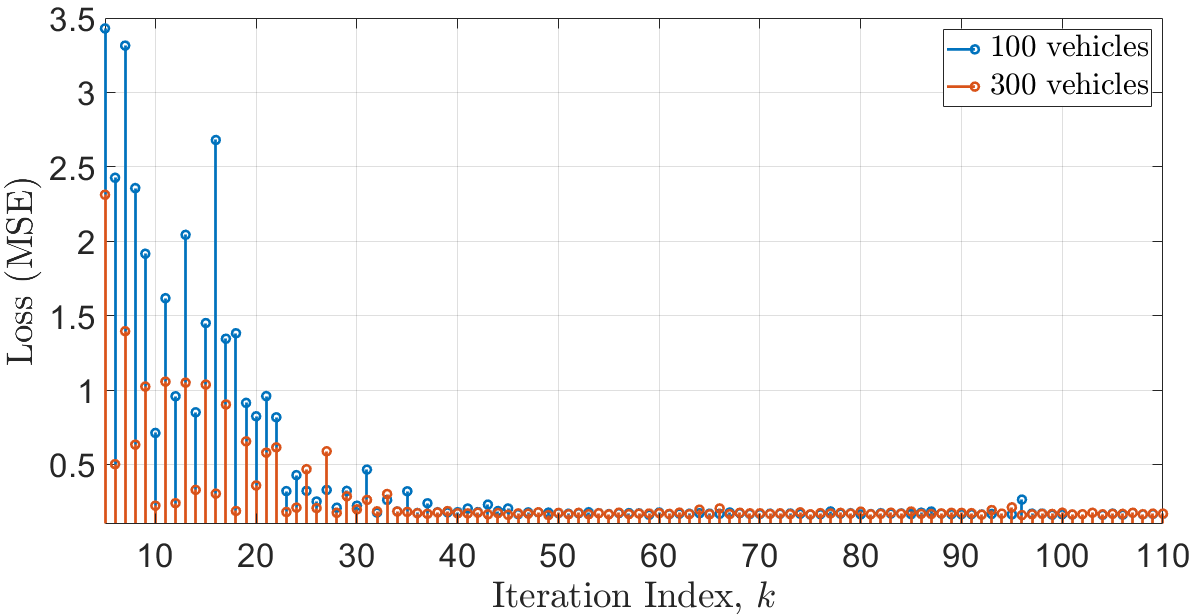}
		\caption{Security check on both dataset and weights.} 
	\end{subfigure}
	\caption{Loss (MSE) of global model with 50\% adversary.} \label{figconv}
\end{figure*}
\begin{figure*}
	\begin{subfigure}{0.3\textwidth}
		\includegraphics*[scale=0.18]{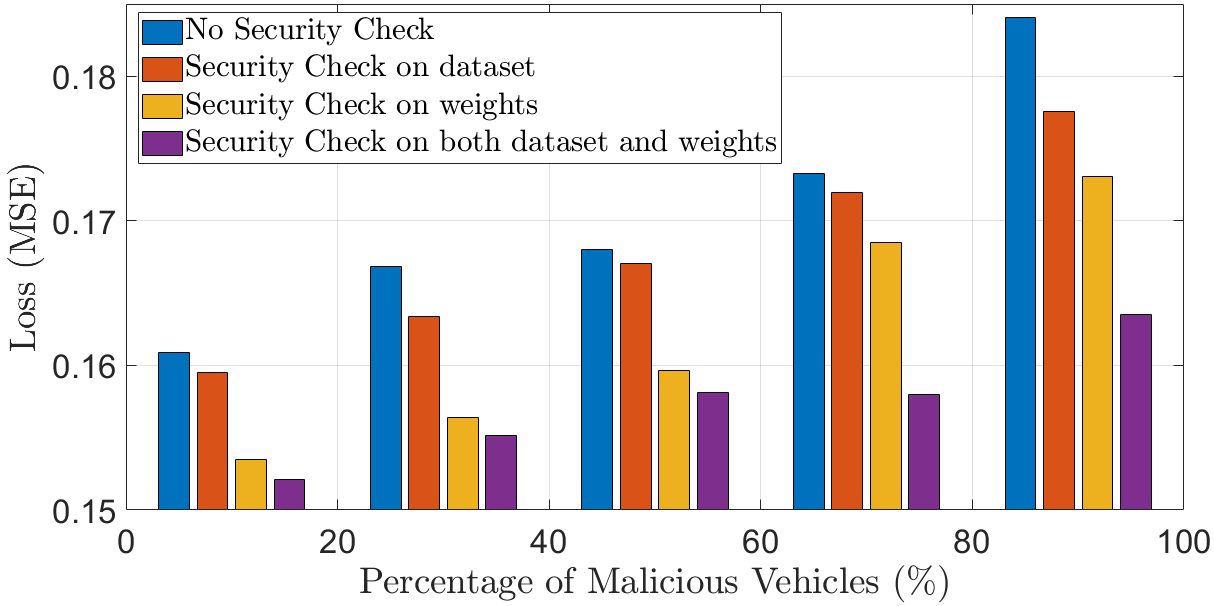}
		\caption{Vehicles: 100, Adversary: Malicious.}
	\end{subfigure}
	\hspace*{\fill} % separation between the subfigures
	\begin{subfigure}{0.3\textwidth}
		\includegraphics*[scale=0.18]{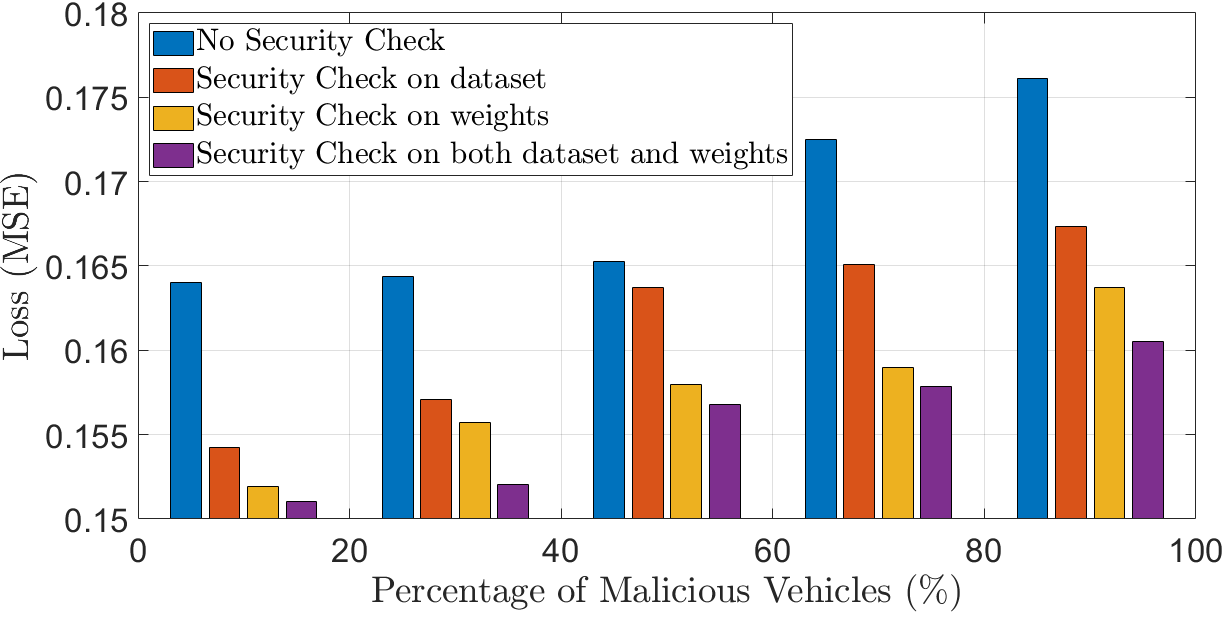}
		\caption{Vehicles: 200, Adversary: Malicious.} 
	\end{subfigure}
	\hspace*{\fill} % separation between the subfigures
	\begin{subfigure}{0.3\textwidth}
		\includegraphics*[scale=0.18]{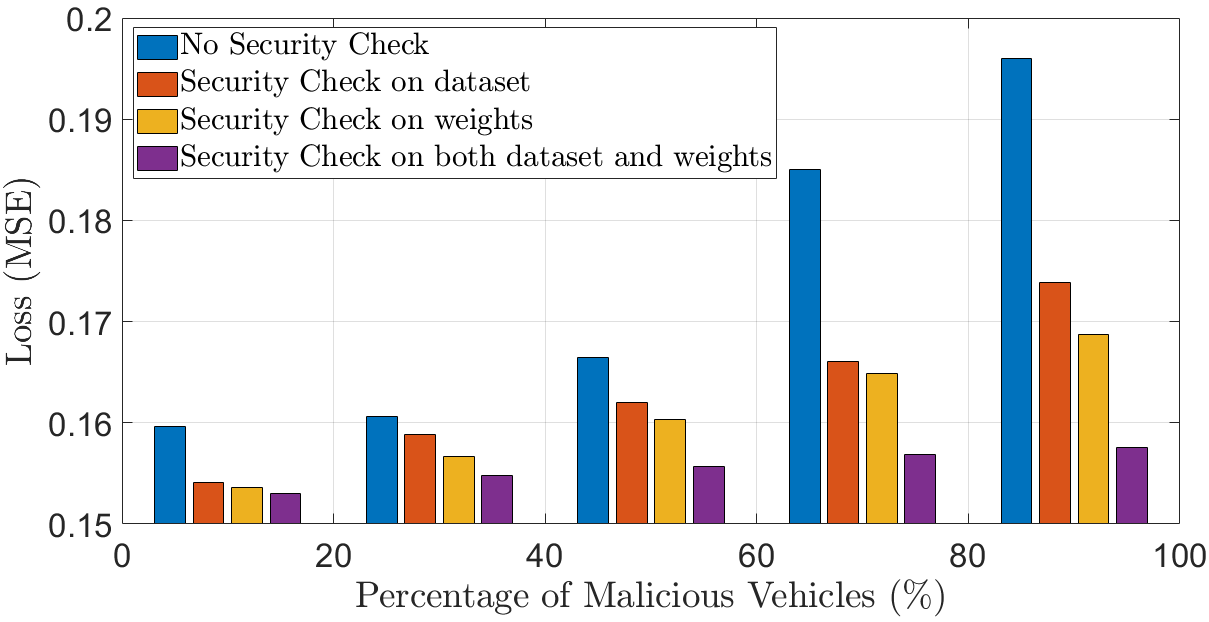}
		\caption{Vehicles: 300, Adversary: Malicious.} 
	\end{subfigure}
%	\hspace*{\fill} % separation between the subfigures
	\begin{subfigure}{0.3\textwidth}
		\includegraphics*[scale=0.18]{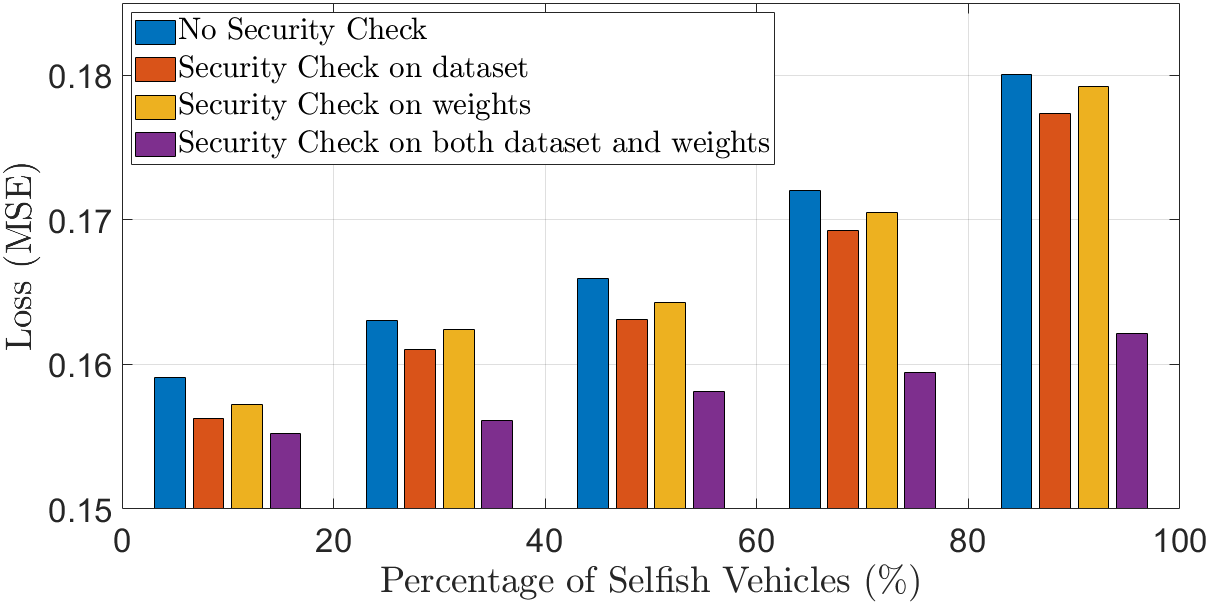}
		\caption{Vehicles: 100, Adversary: Selfish.}
	\end{subfigure}
	\hspace*{\fill} % separation between the subfigures
	\begin{subfigure}{0.3\textwidth}
		\includegraphics*[scale=0.18]{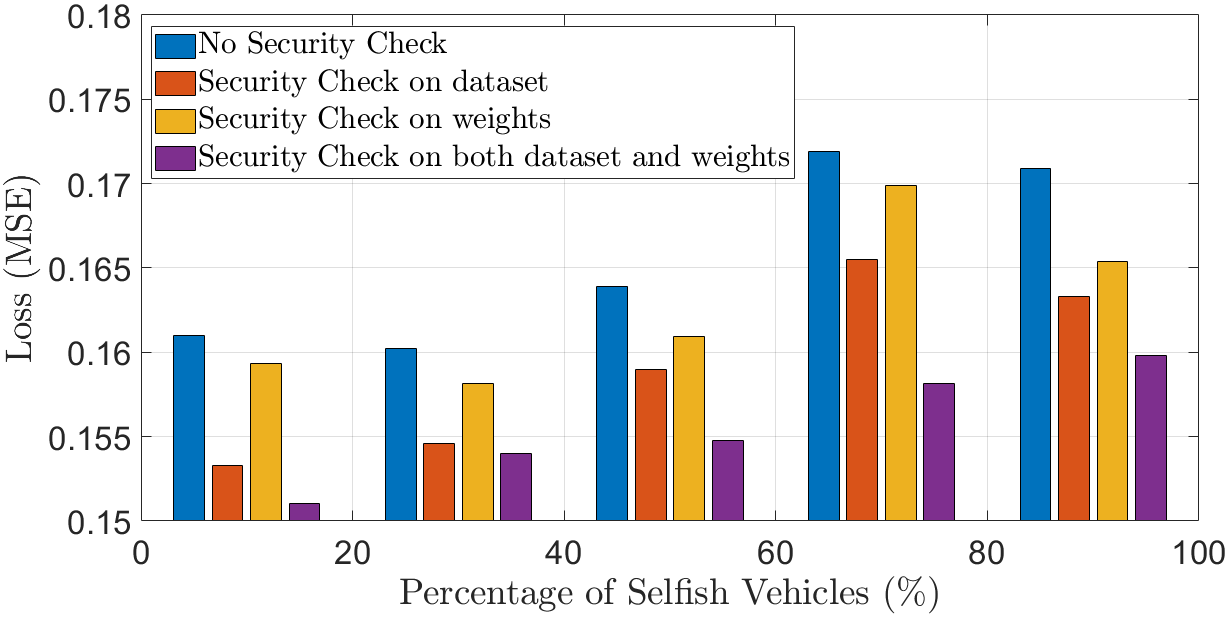}
		\caption{Vehicles: 200, Adversary: Selfish.} 
	\end{subfigure}
	\hspace*{\fill} % separation between the subfigures
	\begin{subfigure}{0.3\textwidth}
		\includegraphics*[scale=0.18]{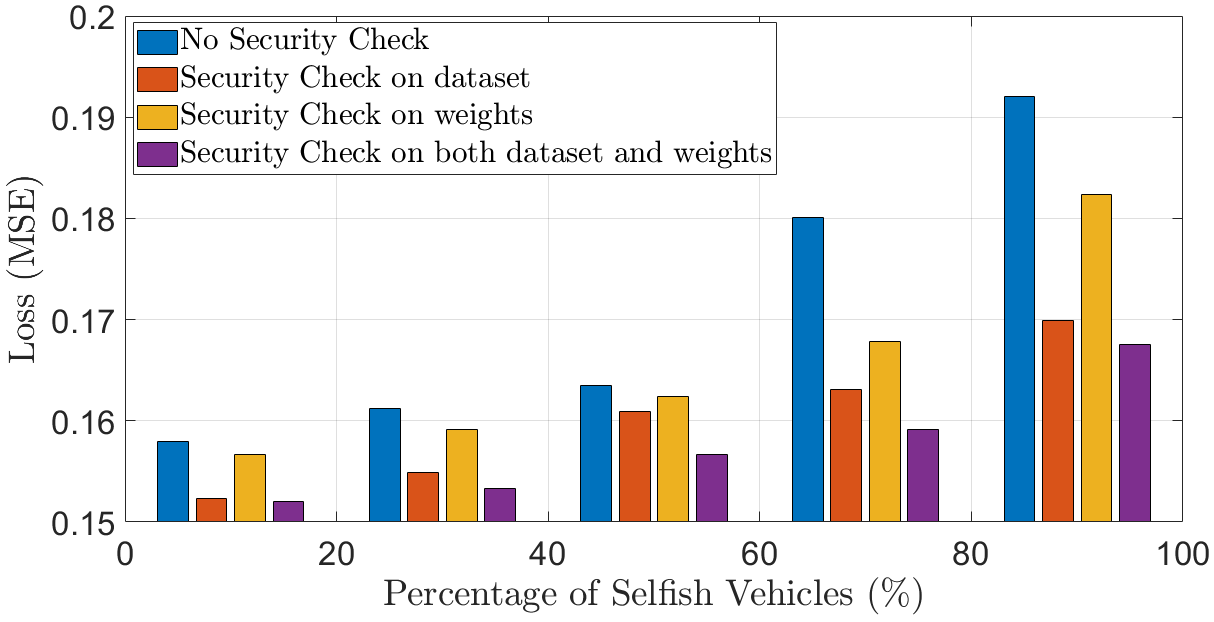}
		\caption{Vehicles: 300, Adversary: Selfish.} 
	\end{subfigure}
	\hspace*{\fill} % separation between the subfigures
	\caption{Loss (MSE) of global model after 100 iterations.} \label{figloss}
\end{figure*}
\begin{figure}
	\centering
	\includegraphics[scale=0.25]{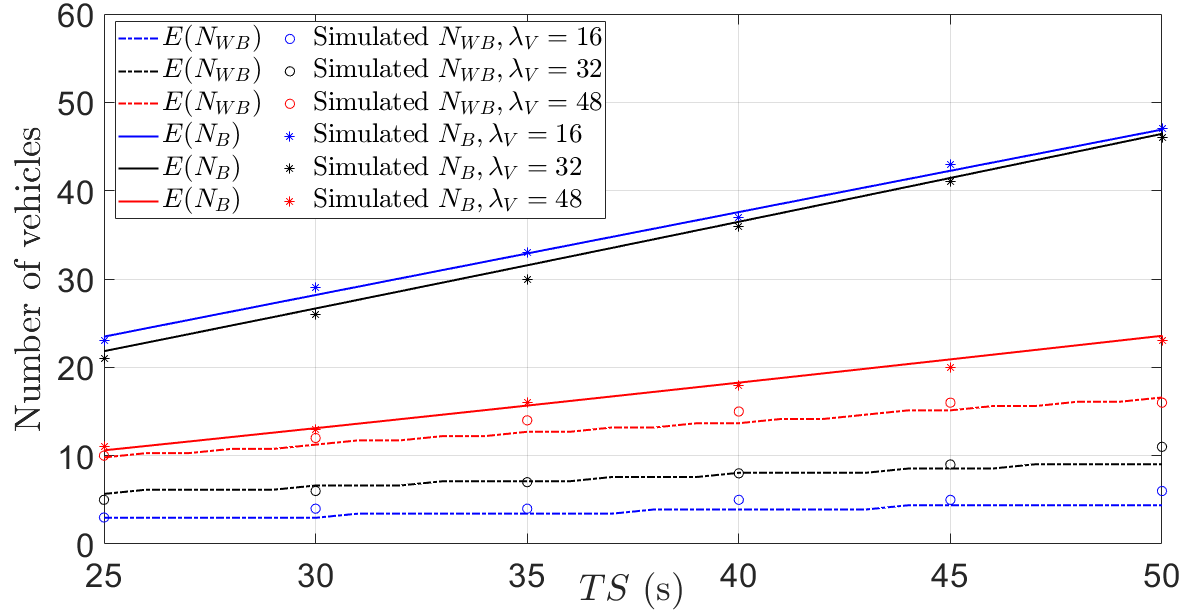}
	\caption{Number of vehicles uploading local model in $TS$.}
	\label{figblock}
\end{figure}
\begin{table}[t]
	\renewcommand{\arraystretch}{1.3}
	\caption{Loss (MSE) of global model after 100 iterations.}
	\label{tableloss}
	\centering
	\begin{tabular}{|c|c|c|c|}
		\hline
		$s_i$ & $N=100$ & $N=200$ & $N=300$ \\
		\hline
		2000 & 0.19643 & 0.18724 & 0.16541 \\
		\hline
		5000 & 0.17251 & 0.17021 & 0.16313 \\
		\hline
		8000 & 0.15297 & 0.15101 & 0.15085 \\
		\hline
	\end{tabular}
\end{table}
\begin{table} [t]
	\renewcommand{\arraystretch}{1.3}
	\caption{$\lambda_{MB}$ and $\mu_d$ with respect to $\lambda_V$.}
	\label{tablelambda}
	\centering
	\begin{tabular}{|c|c|c|c|}
		\hline
		$\lambda_{V}$ (vehicle/m$^2$) & 16 & 32 & 48 \\
		\hline
		$\lambda_{MB}$ (microblocks/s) & 2.01 & 1.99 & 0.98 \\
		\hline
		$\mu_d$ (m) & 344 & 298 & 276 \\
		\hline
	\end{tabular}
\end{table}
\begin{figure*}
	\begin{subfigure}{0.3\textwidth}
		\includegraphics*[scale=0.18]{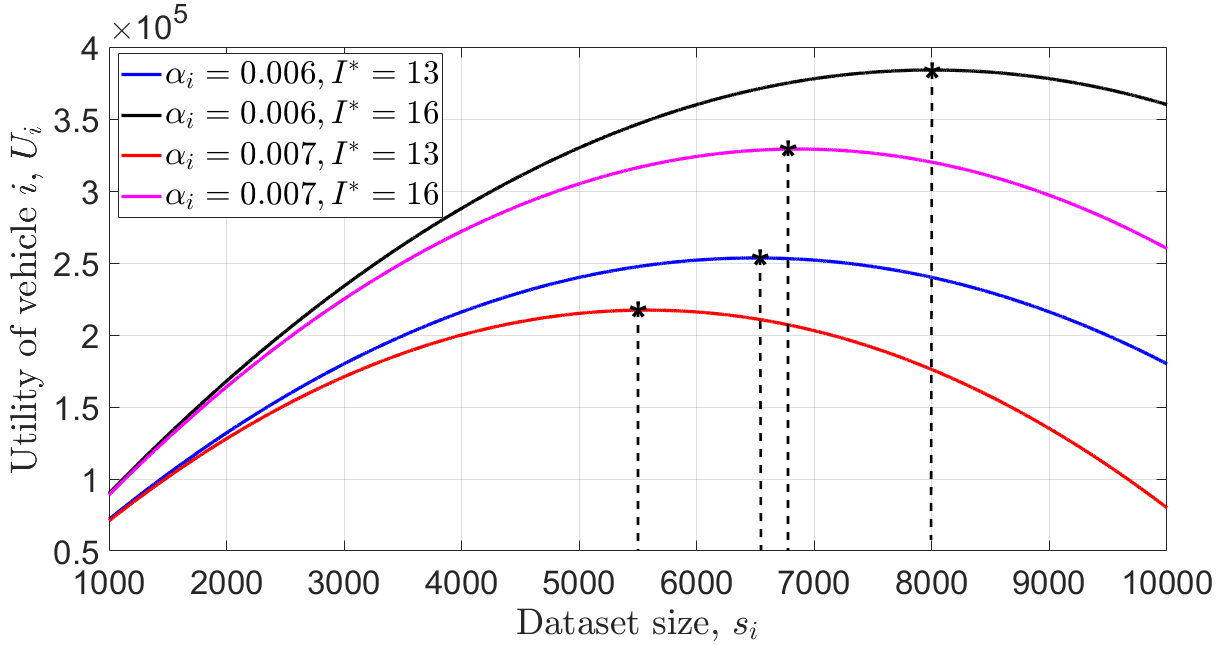}
		\caption{$U_i$.}
	\end{subfigure}
	\hspace*{\fill} % separation between the subfigures
	\begin{subfigure}{0.3\textwidth}
		\includegraphics*[scale=0.18]{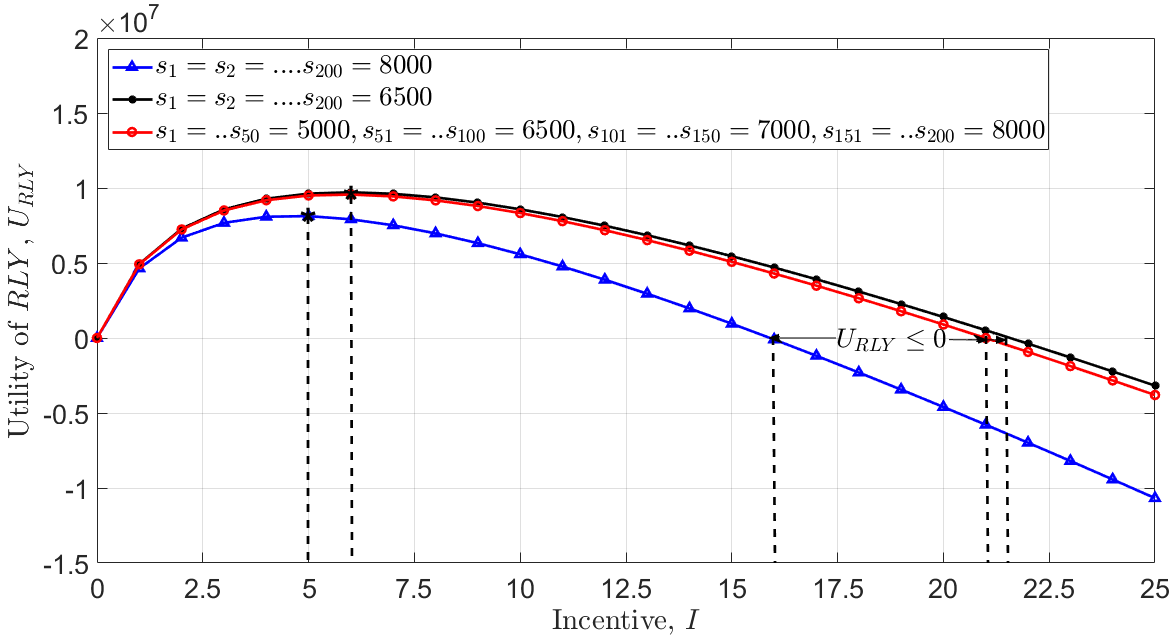}
		\caption{$U_{RLY}$ at $N=200$, $\beta=0.9 \times 10^7$.} 
	\end{subfigure} \hspace*{\fill} 
\begin{subfigure}{0.3\textwidth}
	\includegraphics*[scale=0.18]{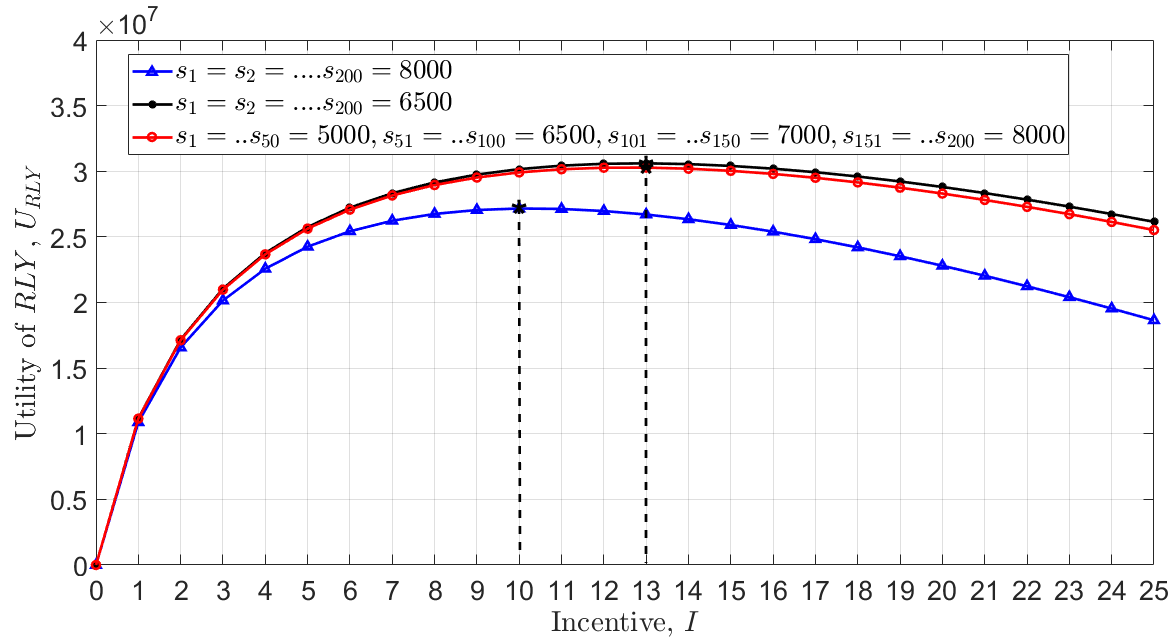}
	\caption{$U_{RLY}$ at $N=200$, $\beta=1.8 \times 10^7$.} 
\end{subfigure}
	\caption{Utility of players in Stackelberg Game with equilibrium points (*).} \label{figStalk}
\end{figure*}
\begin{figure*}
	\begin{subfigure}{0.45\textwidth}
		\includegraphics*[scale=0.24]{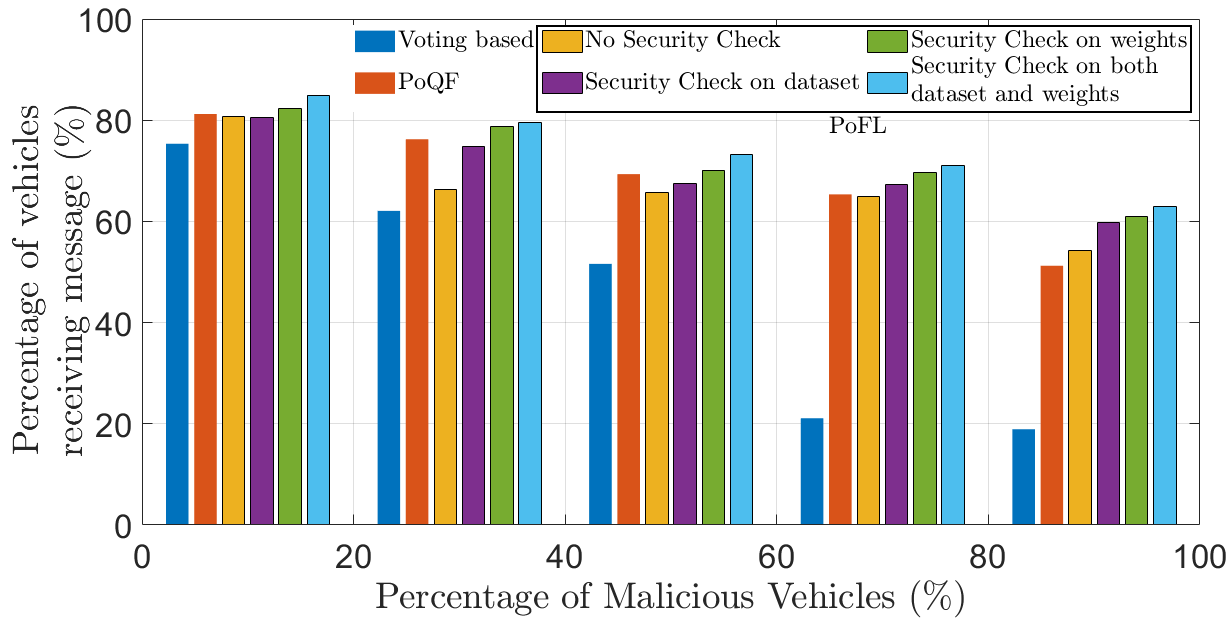}
		\caption{Maximum speed = 55\,km/hr.}
	\end{subfigure}
	\hspace*{\fill} % separation between the subfigures
	\begin{subfigure}{0.45\textwidth}
		\includegraphics*[scale=0.24]{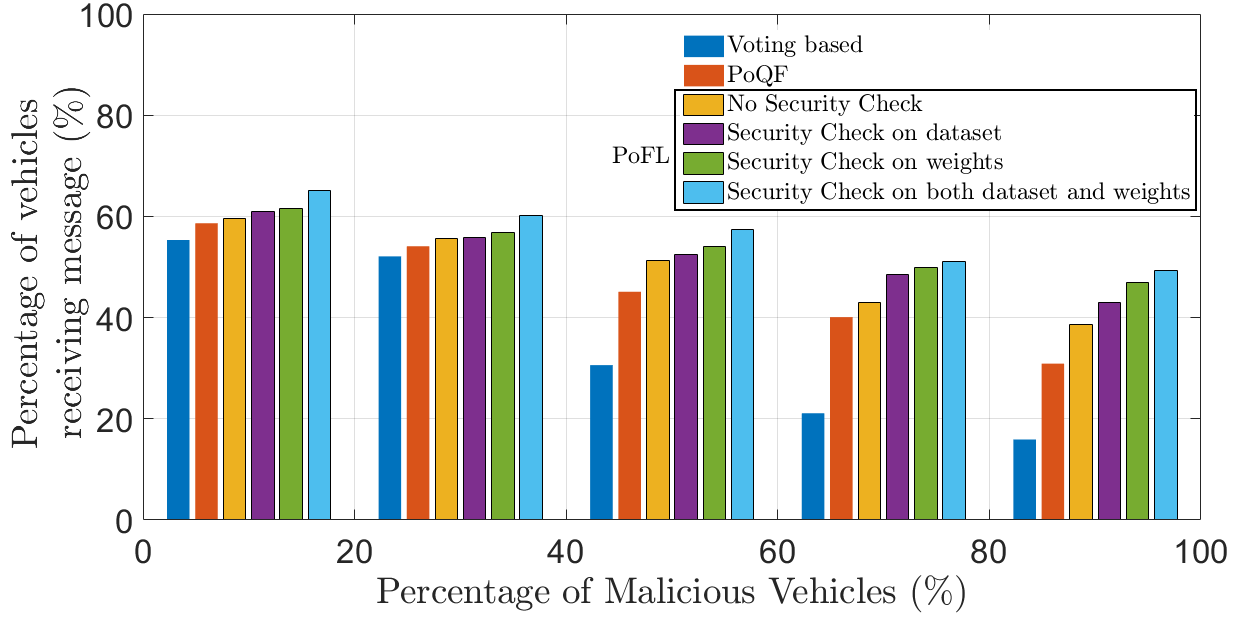}
		\caption{Maximum speed = 110\,km/hr.} 
	\end{subfigure}
	\hspace*{\fill} % separation between the subfigures
	\caption{Message delivery ratio with 300 vehicles.} \label{figrelay}
\end{figure*}
\begin{figure}
		\includegraphics*[scale=0.25]{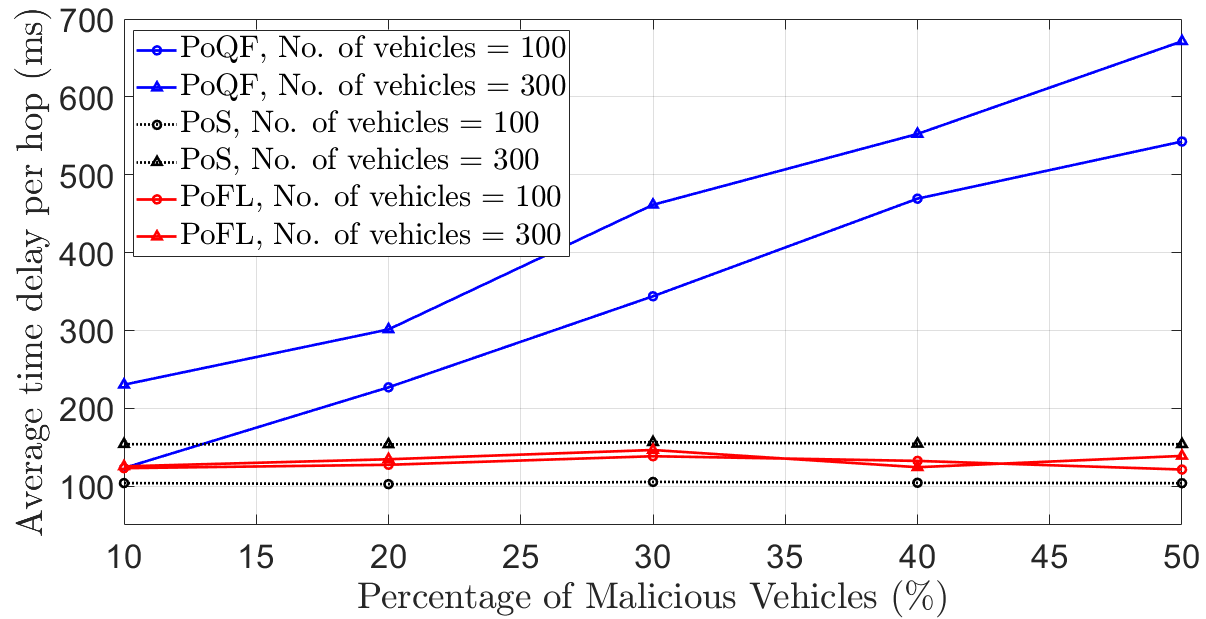}
		\caption{Average time delay per hop.} \label{timedelay}
\end{figure}
Fig.~\ref{figconv} shows the loss (MSE) of global model with respect to iteration index $k$, in presence of 50\% adversary among the vehicles participating in FL, consisting of equal percentage of malicious and selfish vehicles. In all cases, the loss converges to its lowest possible value until 100 iterations. However, this convergence is achieved in less number of iterations with 300 vehicles as compared to 100 vehicles, which means that the maximum accuracy of a global model can be attained faster with greater number of vehicles. Fig.~\ref{figconv}\,(a) shows the loss when no security check is implemented. The convergence rate is slower without security check and takes more iterations than those with security checks, as shown in Fig.~\ref{figconv}\,(b) - (d). It indicates that the increased computation required to implement security check can be compensated with less iterations processed to attain maximum accuracy. Table~\ref{tableloss} shows loss of global model after 100 iterations with respect to number of vehicles participating in FL without any adversary or security check. As shown in Table~\ref{tableloss}, the loss is inversely proportional to both dataset size and number of vehicles. 

Fig.~\ref{figloss} shows the loss of global model after 100 iterations of FL in presence of malicious and selfish vehicles. The global loss function is the highest if no security check is employed in smart contract of FL blockchain. Security check on weights results in less loss as compared to security check on dataset in presence of malicious vehicles and vice-versa in case of selfish vehicles. It is because selfish vehicles affect only $N_A$ in dataset by not sending acknowledgments and such discrepancy is easily detected if security check is applied on dataset only. On the other hand, malicious vehicles can change all parameters in dataset and therefore it is not easy to detect anomaly on such dataset. It shows that security check on weights is more suitable to prevent poisoning attack caused by malicious vehicles and security check on dataset is more appropriate to reduce the effect of selfish behavior. Nevertheless, malicious vehicles may submit a true dataset for security check and upload inaccurate local models using a false dataset. Therefore, in this case, only security check on weights can prevent adversary caused by malicious vehicles. The loss function is minimum if security check on both dataset and weights is used and is suitable for both malicious and selfish vehicles. As a trade-off, increased computation is required to run security check twice. 

Fig.~\ref{figblock} shows the average number of vehicles over 100 simulation runs which uploaded their local models during $TS$, with and without blockchain at various $\lambda_V$. The simulation results are matched with expected values derived in (\ref{eqB}) and (\ref{eqWB}), confirming our theoretical analysis. $\lambda_{MB}$ and $\mu_d$ change with varying $\lambda_V$ and are listed in Table~\ref{tablelambda}. Fig.~\ref{figblock} shows that blockchain based approach results in average 18 more vehicles uploading their local models within same $TS$ compared with the centralized solution in submitting local models directly to RSU without blockchain. This is because a copy of FL blockchain is possessed by each vehicle. A local model by vehicle $i$ can be entered into FL blockchain without depending upon RSU. Subsequently, RSU is able to receive an updated FL blockchain by another vehicle $j$, containing local models of both vehicle $i$ and vehicle $j$. Without blockchain, a vehicle $i$ has to travel towards RSU within $TS$ to directly share its local model. In this case, one RSU or small $TS$ may not be sufficient for receiving local models from large number of vehicles. Also, as shown in Table~\ref{tableloss}, the loss of global model decreases with rising $N$. It can be concluded that FL blockchain can achieve desired accuracy of a global model faster than FL carried out without blockchain, because FL blockchain enables collection of local models from more number of vehicles within the same time limit.

Fig.~\ref{figStalk} proves Definition 1 at equilibrium points at various values of $\alpha_i$, $\beta$ and $N$. Fig.~\ref{figStalk}\,(a) shows the utility of vehicle $i$, $U_i$, participating in a blockchain based FL. As shown in Fig.~\ref{figStalk}\,(a), for a given $I^*$, there exists only one $s_i^*$ which results in maximum $U_i$. Fig.~\ref{figStalk}\,(b) and (c) show $U_{RLY}$ at $N=200$ with varying values of $s_i$ and $\beta$. In each case, an equilibrium point exists where $U_{RLY}$ is maximum for a given $I^*$. A Central Authority can select the value of $I^*$, which gives both maximum $U_i$ and $U_{RLY}$. As shown in Fig.~\ref{figStalk}\,(b), $U_{RLY}<=0$ for certain values of $I$, which will motivate $RLYs$ to become selfish. An appropriate value of $\beta$ can be selected to make $U_{RLY}>0$ for every value of $I$, as shown in Fig.~\ref{figStalk}\,(c). A machine learning model can be used to predict the optimum values of $\beta$ and $I$, according to $s_i$, which result in best response strategy of $RLYs$. This model can be embedded into smart contract of message blockchain to automate reward distribution independently without Central Authority. 

Fig.~\ref{figrelay} shows message delivery ratio among 300 vehicles at varying percentages of malicious vehicles in the network as a result of 100 simulation runs. Results are also compared with PoQF \cite{PoQF} and another voting blockchain based relay selection method in which an appropriate relay is elected on the basis of its distance from the sender and channel quality parameters \cite{ICC}. PoFL outperfoms voting based relay selection method and PoQF when security check is applied. However with low percentage of malicious vehicles and maximum speed of 55\,km/hr, PoQF results in better message delivery ratio than PoFL when security check on weights is not applied. Nevertheless, PoFL with security check on both dataset and weights always outperforms voting based relay selection and PoQF by an average of 25\% and 8.2\% increase in message delivery ratio respectively.

Fig.~\ref{timedelay} shows the average time delay per hop in completing PoFL, PoQF \cite{PoQF} and PoS (Proof-of-Stake) \cite{PoS} consensus. PoS is simulated such that it selects relay node on the basis of reputation of vehicle. A random reputation value following uniform distribution, ranging from 0 to 100 is assigned to each vehicle. The average time delay per hop of PoQF rises with increasing number of vehicles and percentage of malicious vehicles in the network. This is because PoQF waits for a threshold number of votes to determine a relay node and the optimum threshold increases with rising number of vehicles and malicious vehicles percentage. Time delay of PoS rises with increasing number of vehicles due to more time required in accessing large amount of reputation values but it is independent of percentage of malicious vehicles. PoFL is run by each vehicle simultaneously and therefore its time delay is independent of both number of vehicles and percentage of malicious vehicles. On an average, PoFL is 65.2\% faster than PoQF in relay selection and is more suitable for time-critical emergency situations. As a trade-off, PoQF only involves Quality Factor calculations but PoFL is based on a computationally expensive FL process with multiple iterations. Compared to PoS, PoFL is 15.74\% faster when there are 300 vehicles but 18.93\% slower when there are 100 vehicles. This is because PoS consumes time only in accessing the blockchain to find reputation of vehicles. The access time increases when there is a large number of vehicles registered in a blockchain network. Although PoS with 100 vehicles outperforms both PoQF and PoFL, this  faster consensus for block verification and addition cannot be run independently for appropriate relay selection, unlike PoQF and PoFL. \color{black}

\section{Conclusion}
In this paper, we have proposed a decentralized FL based message dissemination, governed by blockchain. The theoretical and practical performance of uploading local models using blockchain is compared with a centralized approach without blockchain. The proposed FL with blockchain can be considered as a faster approach since it results in greater number of uploaded local models within a given time as compared to a solution without blockchain. Smart contract based security checks are proposed to detect adversary, which result in lower MSE in less number of iterations achieved by global model than FL without security check, after 100 iterations. Compared with other blockchain approaches suitable for relay selection in vehicular networks, the proposed solution is 65.2\% faster and at least 8.2\% more efficient in message dissemination approach. It also preserves privacy of neighbour vehicles, unlike other relay selection approaches. An economic model for blockchain based FL is also proposed and analyzed using Stackelberg game to determine optimal data size and incentive which result in the best response strategy of vehicles. Message dissemination and relay selection can further be improved in future work by including cross-layer information in dataset, obtained from physical and MAC layers.

\end{document}